\documentclass[a4paper,fleqn,usenatbib]{mnras}

\usepackage[T1]{fontenc}
\usepackage{ae,aecompl}

\usepackage{graphicx}	
\usepackage{amsmath}	
\usepackage{amssymb}	

\newcommand{\unit}{1\!\!1}

\def\vel{\, \mathbf v}
\def\Bfield{\, \mathbf B}
\def\Efield{\, \mathbf E}
\def\Fpart{\, \mathbf F_{\rm part}}
\def\upart{\, \mathbf u_{\rm part}}
\def\npart{\, n_{\rm part}}
\def\Jpart{\, \mathbf J_{\rm part}}

\title[PIC-MHD models of shocks and particle acceleration]{On magnetic field amplification and particle acceleration near non-relativistic astrophysical shocks:\\ Particles in MHD Cells simulations}

\author[van Marle, Casse \& Marcowith]{
Allard Jan van Marle,$^{1}$\thanks{E-mail: vanmarle@apc.univ-paris7.fr}
Fabien Casse,$^{1}$
Alexandre Marcowith$^{2}$
\\
$^{1}$Laboratoire AstroParticle \& Cosmologie (APC), Universit\'e Paris Diderot, CNRS/IN2P3, CEA/Irfu, Observatoire de Paris,\\
Sorbonne Paris Cit\'e, 10, rue Alice Domon et Leonie Duquet, F-75205 Paris Cedex 13, France.\\
$^{2}$Laboratoire Univers et Particules de Montpellier (LUPM) Universit{\'e} Montpellier, CNRS/IN2P3, CC72, place Eug{\`e}ne Bataillon,\\ 34095, Montpellier Cedex 5, France.\\
}
\date{Accepted XXX. Received YYY; in original form ZZZ}

\pubyear{2017}

\begin{document}
\label{firstpage}
\pagerange{\pageref{firstpage}--\pageref{lastpage}}
\maketitle

\begin{abstract}
We present simulations of magnetized astrophysical shocks taking into account the interplay between the thermal plasma of the shock and supra-thermal particles. 
Such interaction is depicted  by combining a grid-based magneto-hydrodynamics description of the thermal fluid with particle in cell techniques devoted to the dynamics of supra-thermal particles. 
This approach, which incorporates the use of adaptive mesh refinement features, is potentially a key to simulate astrophysical systems on spatial scales that are beyond the reach of pure particle-in-cell simulations. 
We consider in this study non-relativistic shocks with various Alfv\'enic Mach numbers and magnetic field obliquity. We recover all the features of both magnetic field amplification and particle acceleration from previous studies when the magnetic field is parallel to the normal to the shock. 
In contrast with previous particle-in-cell-hybrid simulations, we find that particle acceleration and magnetic field amplification also occur when the magnetic field is oblique to the normal to the shock but on larger timescales than in the parallel case. 
We show that in our simulations, the supra-thermal particles are experiencing acceleration thanks to a pre-heating process of the particle similar to a shock drift acceleration leading to the corrugation of the shock front. 
Such oscillations of the shock front and the magnetic field locally help the particles to enter the upstream region and to initiate a non-resonant streaming instability and finally to induce diffuse particle acceleration.
\end{abstract}




\section{Introduction}
Cosmic Rays (CRs) with energies between a few GeV to hundreds of PeV are produced in our Galaxy. One of the preferred way to accelerate CRs is through repeated crossings of the particles across a shock front carrying magnetic fluctuations responsible for particle scattering \citep{Drury83}. This process is known as diffusive shock acceleration (DSA). It appears that energetic particles (shorten as CRs hereafter) can be accelerated so efficiently that they can drive their own magnetic fluctuations and sustain the acceleration process \citet{Bell78}. DSA can become so efficient that CR pressure modifies the shock structure and the acceleration process requiring non-linear CR back-reaction calculations to derive self-consistent solutions \citep{Berezhko99}. It appears that CRs are also able to generate intense magnetic fields at fast shock fronts. This may provide an explanation for the high magnetic field strengths deduced from high-angular resolution observations in the X-ray domain in several historical 
supernova remnants (SNRs) \citep{Parizot06}. In fast shocks that prevail in young SNRs the streaming of CRs ahead the shock front triggers a class of plasma instabilities that can grow fast enough to produce magnetic field strengths sometimes in excess of the standard interstellar values \citep{ZacharyCohen:1986, Lucek00, Bell04} by several orders of magnitude. In particular, \citet{Bell04, Bell05} show that the fastest instability is induced by the return background plasma current which compensates the current produced by streaming CRs. Contrary to its kinetic equivalent, this instability is non-resonant and can be treated using ideal magneto-hydrodynamics (MHD). The non-resonant streaming (NRS) instability has been tested in a long series of numerical studies using various techniques going from pure MHD \citep{Zirakashvili08}, di-hybrid (CRs and protons are treated by particle-in-cell or PIC methods and electrons as a fluid) \citep{Riquelme10, Caprioli14a} to Vlasov  or PIC-MHD \citep{Reville08, Reville13, 
Baietal:2015}; see \citet{Marcowith16} for a review. The study by \citet{Baietal:2015} combines a kinetic description of CRs based on PIC methods coupled to a MHD code which calculate the background magnetized fluid distribution. 

\citet{Reville12, Reville13} used 2D and then 3D simulations to investigate a CR driven filamentation instability which also results from the onset of upstream CR streaming but contrary to the NRS generates long-wavelength perturbations necessary to confine high-energy CRs at the shock front. In particular \citet{Reville13} developed a formalism based on a spherical harmonic expansion of the Vlasov-Fokker-Planck equation which allowed them to investigate 3D configurations. The authors confirmed 2D linear results obtained in \citet{Reville12} and found a kind of universal behavior of the shock precursor: oblique and perpendicular shocks behave very similarly to parallel shocks nearby the shock front where the magnetic field is found completely disordered. 
\citet{Caprioli13} also investigated CR driven filamentation instabilities in a parallel shock configuration but using a di-hybrid (hereafter termed as hybrid) method where electrons are treated as a fluid and protons as kinetic particles. Their 2D and 3D results confirmed the conclusions raised by \citet{Reville12} and show strong magnetic field amplification by NRS instability and further on by filamentation that lead to particle acceleration. \citet{Baietal:2015} have complemented the latter study by simulating parallel shock acceleration for configurations that prevail in young SNR using a PIC-MHD method. The authors have found a fast growth of magnetic field fluctuations in agreement with the linear analysis proposed in \citet{Bell04}. They found that, in the precursor structure, cavitation and filamentation progress over time to larger sizes. They also found evidence of particle acceleration as result of multiple shock crossings. Hybrid simulations  have also been performed in the oblique shock 
configuration by \citet{Caprioli14a} using 2D hybrid simulations. They reported on the drop of acceleration efficiency (defined as the number of particle reaching energies larger than ten times the injection energy) and magnetic field amplification for oblique shocks with angles between the background magnetic field and the shock normal larger than $\sim 50^o$. This result is in apparent conflict  with the results derived by \citet{Reville13}. This issue is addressed in the present study.

In this study we consider the case of non-relativistic shocks corresponding with physical conditions that prevail in SNRs. We investigate several set-ups: parallel and oblique magnetic field configurations, different values of shock velocity and shock Alfv\'enic Mach numbers corresponding to different medium magnetization. Our simulations include the whole system including the shock front whereas \citet{Reville12, Reville13} only included a CR precursor in their analysis. We select the PIC-MHD method which allows us to study large space and time scales of the shock problem whereas the scales included in \citet{Caprioli13} were limited due to the hybrid approach. Finally, we generalize the solutions obtained by \citet{Baietal:2015} as our analysis is not restricted to parallel shocks while displaying for the first time the coupling between an {\it adaptive mesh refinement (AMR)} MHD description to a PIC approach of supra-thermal particle dynamics. AMR grid is likely to become a key element of forthcoming 
numerical simulations if one hopes to describe CR acceleration and transport on the scale of actual astrophysical objects.\\

The paper lay-out is as follows:
\begin{itemize}
\item Section \ref{sec-physics}: Describes the physical model that forms the basis of our treatment of the combined MHD and particle physics as well as a description of our numerical framework.
\item Section \ref{sec-parallel} and section \ref{sec-Bfieldangle}: Show the results of Particle in MHD cells simulations (hereafter PI[MHD]C) dealing with parallel and oblique magnetic field amplification and particle acceleration occurring in the vicinity of non-relativistic astrophysical shocks.
\item Section \ref{sec-discussion}: Discusses of our results with respect to previous studies. 
\item Section \ref{sec-conclusions}: Delivers our conclusions regarding the use of AMR MHD coupled to PIC in order to describe the interplay between the magnetic field, the thermal plasma and the supra-thermal particles.
\end{itemize}

\section{The PI[MHD]C framework}
\label{sec-physics}
\subsection{Physical context}
Over the course of the last half-century, several methods have been developed to produce numerical simulations of the processes involved in the interaction between gas and magnetic field. Two of the most commonly used methods are grid-based MHD and PIC. Whereas the former deals primarily with large-scale structures, at the expense of the micro-physics involved in the interactions, the latter focuses exclusively on the micro-physics. Unfortunately, a large gap in scales exists in between these two regimes, where occur relevant astrophysical interactions for the problem of relativistic particle acceleration which are too large to be modeled effectively using the PIC method, and yet require the microphysics that MHD cannot provide. In this study we use a method which aims at combining these two approaches. This method will henceforth be referred to as Particles in MHD Cells: PI[MHD]C.

\subsection{Grid-based MHD vs Particle-In-Cell}
Grid-based MHD is a tried and proven method for simulating large-scale structures of gas interactions in a magnetic field. Grid-based MHD codes typically solve the equations of continuity and conservation of momentum, as well as an equation that describes the evolution of the energy (which can be conservation of total energy, but other options exist). 
Grid-based MHD is an efficient method that allows to quickly solve large scale problems. Unfortunately, it only deals with statistical properties and does not include particle-based effects. 

Unlike grid-based MHD, the PIC approach to gas dynamics treats the gas as a large collection of particles. This allows the user to describe physical effects that are based on the behavior of individual particles, something that is not possible in grid-based MHD. The downside of the PIC approach is that it is numerically expensive, has a high level of numerical noise, and hence can only deal with relatively small physical volumes and/or short physical timescales. Both approaches however share common features as the fact that both have to provide a temporal description of the electromagnetic field occurring within the plasma. PIC codes usually use a Yee-type algorithm to solve the Maxwell equations while MHD code solve the magnetic induction equation. In the non-relativistic MHD limit, it is generally assumed that the inertia of the thermal electrons can be neglected leading to a so-called Ohm's law linking the electric field to the magnetic field so that only magnetic terms enter the various MHD equations. In 
the context of astrophysical shocks where supra-thermal particles carry electrical charges, such relationship between electric and magnetic field has to be revisited.

\subsection{Combining both methods using Ohm's law}
\label{sec-pimhdcphysics}
 The large-scale description of particle acceleration occurring near astrophysical shocks stands as a challenge for PIC codes as such numerical method is inherently computationally expensive if one considers large volume of gas. In many practical applications the majority of the particles behave as a thermal fluid and is described far more efficiently by grid-based MHD. It is only the (relatively) small number of non-thermal particles that need to be treated individually. 
Therefore, the key to handle simultaneously both the thermal fluid and the supra-thermal particles is to achieve a mutual feedback in order to provide a self-consistent description of the interplay between both populations. Such link lies in the electromagnetic field whose interaction with both type of particles is the cornerstone of collisionless plasmas. In the context of non-relativistic MHD, the electric field can be expressed as a function of the magnetic field and the fluid velocity via a so-called Ohm's law. The presence of supra-thermal particles within the thermal plasma leads to the addition of extra terms within the Ohm's law taking into account the electrical current and charge carried by these particles.
Such an approach has been presented by \citet{Baietal:2015} considering a single particle species with a positive electrical charge. In the next subsection we generalized the aforementioned approach to any type of supra-thermal particles (electrons and ions). 

\subsubsection{PI[MHD]C equations}
The MHD conservation equations for mass, momentum and energy, including the additional terms that arise from the interaction with the non-thermal particles are:
\begin{equation}
\frac{\partial \rho}{\partial t} ~+~ \nabla \cdot (\rho \vel)~=~0,
\label{eq:mass}
\end{equation}
where $\rho$ and $\vel$ stand as the mass density and velocity of the thermal plasma,
\begin{equation}
\frac{\partial \rho\vel}{\partial t} + \nabla\cdot\left(\rho\vel\otimes\vel-\frac{\Bfield\otimes\Bfield}{4\pi}+P_{\mathrm tot}\unit\right)~=~-\Fpart,
\label{eq:momentum}
\end{equation}
with $\Bfield$ being the magnetic field while $P_\mathrm{tot}=P+B^2/8\pi$ is the total pressure. The energy equations reads
\begin{equation}
\frac{\partial e}{\partial t}+\nabla\cdot\biggl( (e + P_{\rm tot})\vel+(\Efield-{\mathbf E}_0)\times\frac{\Bfield}{4\pi}\biggr) 
~=~ -\upart \cdot \Fpart
\label{eq:energy}
\end{equation}
where $e$ is the total energy density of the thermal plasma. The new terms appearing on the rhs of Eq.(\ref{eq:momentum},\ref{eq:energy}) involve an averaged supra-thermal particle velocity
$\upart$ as well as the opposite of the force density applied by the thermal plasma upon the supra-thermal particle $\Fpart$. The definition of $\upart$  stems from the generalized Ohm's law expression, namely
\begin{equation}
c\Efield = -c\frac{\nabla P_e}{n_ee}+\left({\mathbf J}_{\mathrm tot}-n_i e\vel-\sum_{\alpha} n_{\alpha}q_{\alpha}{\mathbf u}_{\alpha}\right)\times\frac{\Bfield}{n_ee}
\end{equation}
where $n_e$ is the number density of thermal electrons, $n_i$ is the number density of thermal ions. $n_{\alpha}$ is the average number density of supra-thermal particle species $\alpha$, $q_\alpha$ being its individual electrical charge while ${\mathbf u}_\alpha$ stands for its average velocity. $e$ is the positive elementary charge. We note the total non-thermal density as $n_{\rm part} = \sum_{\alpha} n_{\alpha}q_{\alpha}/e$. From charge conservation we have $n_{\rm e}= n_{\rm i}+ n_{\rm part}$. As discussed by \citet{Baietal:2015}, the terms related to the electron thermal pressure and the Hall current can be safely discarded provided we consider plasmas where the magnetic field pressure is not very small compared to thermal pressure and typical variation lengthscales larger than the ion skin depth. Considering the local electrical neutrality of the plasma and supra-thermal population altogether, we obtain the expression of Ohm's law, namely
\begin{equation}\label{Eq:ELE}
c\Efield = -\left((1-R)\vel +R\upart\right)\times\Bfield
\end{equation}
where $R=\sum_{\alpha} n_{\alpha}q_{\alpha}/n_ee$ is a direct measure of the relative density of supra-thermal particles. The average velocity of the whole supra-thermal population is 
\begin{equation}
\upart=\frac{\sum_{\alpha} n_{\alpha}q_{\alpha}{\mathbf u}_{\alpha}}{\sum_{\alpha} n_{\alpha}q_{\alpha}}=\frac{\Jpart}{\npart e}
\end{equation}
Additionally we define
\begin{equation}
{\mathbf E}_0~=~ -\frac{\vel}{c}\times\Bfield
\end{equation}
as the electric field that would be generated by the thermal gas alone. The actual electric field, $\Efield$, generated by both thermal and non-thermal populations combined is related to the force applied by the thermal gas upon supra-thermal particles, namely
\begin{equation}
\frac{\Fpart}{n_i e}~=~(1-R)\left( {\mathbf E}_0-\Efield\right)
\end{equation}
The Lorentz force $\Fpart$ can be expressed as
\begin{equation}
\Fpart~=~(1-R) \biggl( \npart e {\mathbf E}_0 + \frac{\Jpart}{c} \times \Bfield \biggl), 
\end{equation}
In order to close the set of PI[MHD]C equations, we consider the temporal evolution of the magnetic field provided by the Maxwell-Ampere equation,
\begin{equation}
\frac{\partial\Bfield}{\partial t}~=~c\nabla \times \Efield
\end{equation}
For the sake of completeness we have to consider the total electric field generated by both populations  in the induction equation. 
\subsubsection{Particle motion}
Most of the astrophysical shocks involve collisionless plasmas where one can safely discard collisions between particles. The equation governing the motion of individual supra-thermal particles will then only take into account the electromagnetic force, namely 
\begin{equation}
\frac{\partial \mathbf{p}_{\alpha ,j}}{\partial t}~=~q_j\biggl(\Efield+ \frac{{\mathbf u}_{\alpha,j}}{c}\times\Bfield \biggr)
\end{equation}
with ${\mathbf p}_{\alpha ,j}$, $q_j$ and ${\mathbf u}_{\alpha ,j}$ the momentum, charge and velocity of particle $j$. Inserting Eq.(\ref{Eq:ELE}) for the electric field leads to the following equation 
\begin{equation}
\frac{\partial \mathbf{p}_{\alpha,j}}{\partial t}~=~\frac{q_{\alpha}}{c}\left({\mathbf u}_{\alpha,j}-(1-R)\vel-R\upart\right)\times\Bfield
\end{equation}
where we see that contributions from the thermal plasma and the average velocity of supra-thermal particle do influence the motion of a single particle. Within a given MHD cell, one can express the energy balance regarding all supra-thermal particles located in this cell as 
\begin{equation}
\sum_{\alpha}\sum_jn_{\alpha}{\mathbf u}_{\alpha,j}\cdot\frac{\partial \mathbf{p}_{\alpha,j}}{\partial t}=\upart\cdot\Fpart
\end{equation}
which shows that the total energy is conserved as the supra-thermal population exchanges energy with the thermal plasma and magnetic field. 
\subsection{Numerical approach}
\label{sec-numerics}
We use the {\tt MPI-AMRVAC} \citep{vanderHolstetal:2008} code as a basis. This is a fully conservative, finite-volume code that solves conservation equations on an adaptive mesh grid and uses the OpenMPI library for running in parallel on systems with distributed memory architecture. In order to create a high-performance PI[MHD]C code, we maintain the existing architecture of the code but replace the standard equations of ideal magneto-hydrodynamics with Eqs.~\ref{eq:mass}-\ref{eq:energy}. We have also created an additional module which includes the additional physics required to deal with PIC treatment of individual particles.

The {\tt MPI-AMRVAC} code is fully 3-D, but for the study presented in this paper, we limit ourselves to 2D3V applications, namely using a two-dimensional spatial grid while velocity and electromagnetic field exhibit three components. In the following paragraphs we present the various new features added to the {\tt MPI-AMRVAC} code.

\subsubsection{PI[MHD]C Code structure}
Finite volume MHD codes rely on a grid of given geometry where MHD quantities are defined at the center of each cell of the grid. In order to compute the MHD fluxes occurring through each cell interface, one needs to accurately approach the solution of the corresponding Riemann problem using a given MHD solver combined to a slope limiter. On the other hand, PIC codes usually define physical quantities at different locations on the grid when they employ a Yee-type algorithm. For instance, magnetic fields are defined at the center of the cells while electric fields and current density are set on the cell edges and charge density at the corners.

In order to couple both approaches with the minimum amount of numerical computations, we choose to use the MHD grid as a base and to consider an offset PIC grid where MHD cell centers stand as PIC cell corners. 
Such choice allows us to simply map the charge and current densities generated by the particles onto the MHD grid without any interpolation procedure. Mapping the charge and current densities is achieved through standard second-order cloud-in-cell techniques \citep{FerrellBertschinger1994}.

The temporal evolution of the simulation follows the following pattern: At the beginning of each time-step, the MHD quantities are saved. We then advance them for one MHD time-step according to the equations derived in Section~\ref{sec-pimhdcphysics}, which include both the magneto-hydrodynamical properties of the thermal gas and the charge and current generated by the non-thermal particles.
We then evolve the particle positions and velocities of the supra-thermal particles using a relativistic form of the Boris-method \citep[][and references therein]{BirdsallLangdon:1991} based on the MHD quantities from the beginning of the MHD time-step. Finally, we map the charge and current densities, as determined by the new particle distribution in phase space, onto the grid. At this point the time-step ends and the process repeats itself. It is noteworthy that MHD time-step (based on Courant-Friedrich-Levy condition) and PIC time-step (imposed by particle dynamics) do not match. As PIC time-step is usually smaller than the MHD ones, we have the possibility to perform several PIC time-step within one given MHD step. However, in order to maintain the simulation coherence, we do not allow more than a few PIC steps within one MHD step (see \citet{Baietal:2015}). 

\begin{figure*}
\centering
\mbox{
\includegraphics[width=0.3\textwidth]{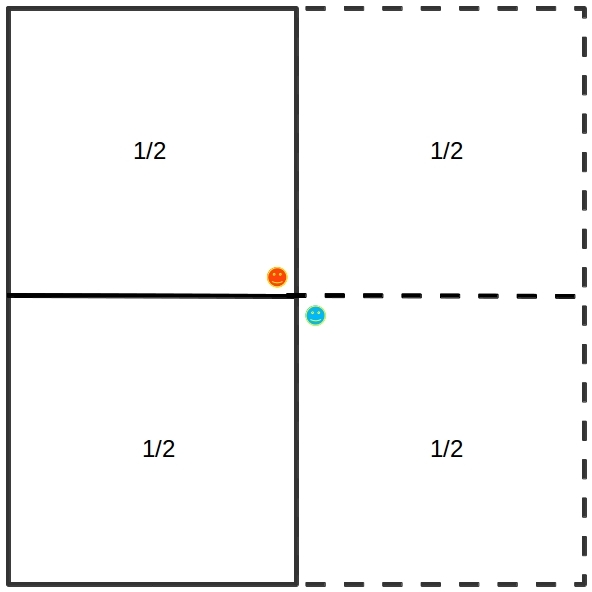}
\includegraphics[width=0.3\textwidth]{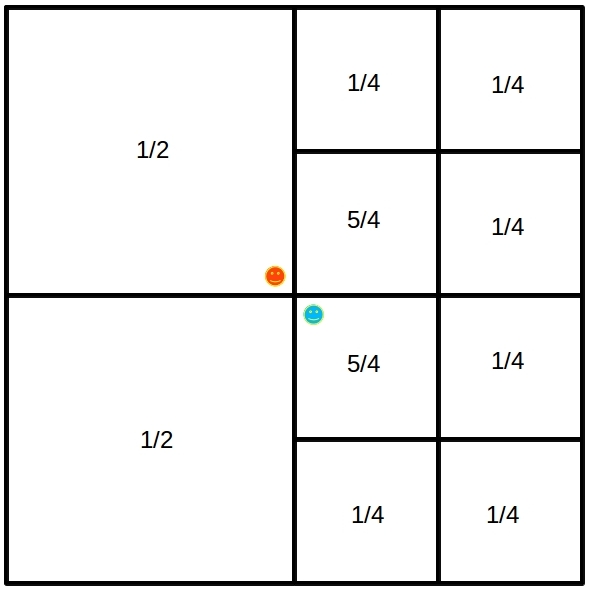}
\includegraphics[width=0.3\textwidth]{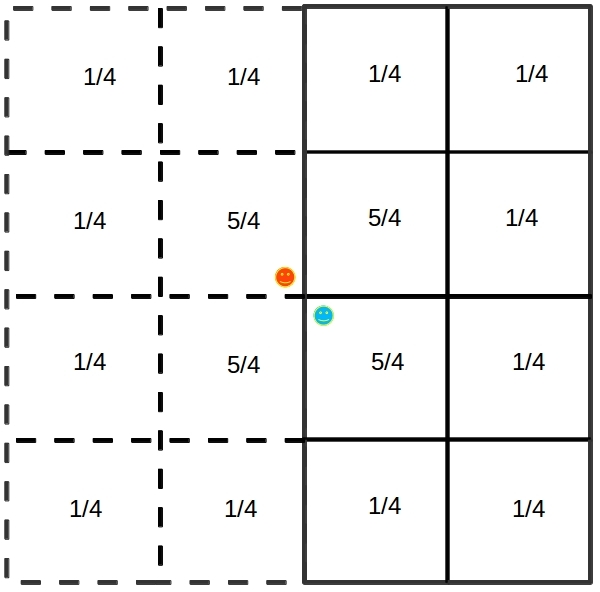}}
\caption{Distribution of particle weights at the edge of two grids with different levels of refinement. Two particles are sitting close together near the intersection. One (red) just inside the coarser grid, the second (blue) just inside the finer grid. Continuous lines show the actual domain, the dashed lines are ghostcells.  
The left-side figure shows how it is perceived by the coarser grid, the right-side figure shows how it is perceived by the finer grid. In both cases, the charge distribution is symmetrical around the two particles. 
The figure at the center shows the actual charge density as seen in the output. 
}
 \label{fig:mesh}
\end{figure*}

\subsubsection{Adaptive Mesh Refinement}
The {\tt MPI-AMRVAC} code uses the Octree system of adaptive mesh refinement \citep{ShephardGeorges:1991}. This system divides the physical domain of the simulation into a number of identical grids, which can be independently refined. Each time a grid is refined, it is subdivided into $2^D$ similar grids that are $D$-dimensional. The decision on whether or not to refine is based on the variance of one or more selected variables  within a grid. If the variance exceeds a pre-set limit then the grid is refined. Conversely, if the variance gets below a given limit within a part of the computational domain, then the grid is coarsened. 
The Octree system further ensures that adjacent grids never differ by more than one level of refinement.
Implementing PI[MHD]C, we maintain this structure, but put in an additional condition: if the number of particles within a grid reaches $25\%$ of a pre-set maximum the grid is no longer allowed to coarsen. If the number of particles reaches $80\%$ of the maximum, the grid is refined (assuming it has not yet reached the maximum refinement allowed). 
Additionally, the grid can be set to refine if the average Larmor radius of the particles within the grid becomes smaller than a pre-set number times the size of the individual grid cells. This ensures that the Larmor radius is always resolved (see also \S~\ref{sec-resolution}.) 

At the boundaries between grids, each grid has a set of ``ghostcells'', which provide it with information from its neighbors. For usual MHD without AMR the values in these ghostcells are a simple matter of copying the values from the neighbor. In case AMR is triggered, then one must rely on coarsening/interpolating procedures to provide the necessary data from coarser/finer neighboring grids. 
However, PIC-related variables (charge and current density) are a different matter. Because each particle contributes to the surrounding grid cells, a charged particle lying near the boundary will also contribute to the ghostcells and its contribution must be preserved. So, rather than overwriting the contents of the ghostcells with values from neighboring grids, these values have to be added up. Similarly, the original values of the neighbor ghostcells have to be added to the grid cells inside the boundary. This discrepancy between MHD and PIC approaches requires a specific boundary treatment regarding the different quantities while preserving computational efficiency. 

A particular problem that has to be considered when combining PIC treatment with an adaptive mesh is how to deal with particles that move from a coarser grid to a finer one, or the other way round. The area of influence of a particle is determined by the size of the mesh, not the physical domain. As a result, when a particle moves from a coarse to a fine grid, its area of influence is reduced. To compensate for this, and conserve charge and current, the effective weight of the particle has to be increased by a factor equivalent to the reduction in effective volume. 
An alternative method would be to split particles when they enter a more refined grid. However, this would cause additional problems: The number of particles would increase rapidly, negating part of the advantage of the PI[MHD]C method. It would also require additional treatment when the particles returned to a coarser grid: Either we would have to recombine particles (not a practical approach because there is no guarantee that the correct number of particles would leave the more refined area at the same time), or we would have to accept that we would have multiple particle species in each grid, with their charge and mass determined by whether or not a particle had passed through a highly refined area at any given time during the simulation. 
In order to avoid these issues, we do not split particles, but instead scale its effective weight to conserve the charge.

Another concern, when incorporating adaptive mesh refinement and PIC in the same code, is the treatment of boundaries between grids with different levels of resolution. Because the area of influence of each particle is determined by the mesh-size, rather than a physical domain, the boundaries require careful treatment in order to ensure that charge and current density are conserved across the grid boundaries. We achieve this by keeping the physical size of the stencil for particles in the grid with the coarser mesh the same, which means that it gets spread over two rows of gridcells in the  more refined grid (see Fig.~\ref{fig:mesh}). 
This method guarantees that the charge distribution is symmetrical around the particle, both according to the coarser and the finer grid, even though it seems asymmetric in the output.

\subsubsection{Maintaining a divergence free magnetic field}
A potential downside of the finite-volume approach is that it does not guarantee that the magnetic field remains divergence free. This is of particular concern when using the PI[MHD]C method, especially in situations where the number of particles in a given area is low. In order to solve this potential problem we have implemented within the {\tt MPI-AMRVAC} code a Constrained Transport algorithm based on \citet{BalsaraSpicer:1999}. In such approach, numerical fluxes provided by the MHD solver (which also include terms from supra-thermal particles) are used to enforce the solenoid nature of the magnetic field.

\subsubsection{Resolution and time-step control}
\label{sec-resolution}
The initial MHD grid resolution is determined by the user as well as the number of refinement levels allowed in the simulation. Such choice must be done carefully since MHD and PIC methods use fundamentally different criteria to determine the correct resolution. For grid-based MHD the primary consideration for the size of individual grid cells is that they should be smaller than the size of the structures that occur in the simulation, in order to ensure that the morphology of the gas is fully resolved. It is important to recall here that validity of the MHD description imposes that the size of the phenomenon occurring in the simulation has to remain larger than ion skin depth $c/\omega_{\rm p,i}$ where $\omega_{\rm p,i}^2=4\pi n_i e^2/m_i$. In order to enforce that condition, we then make sure that the size of cell embedded in the most refined level of the AMR is larger than the local ion skin depth.

Regarding PIC simulation the required resolution depends on the Larmor radius of the particles, which needs to be well resolved to ensure that the code captures the particle trajectories properly. 
For PI[MHD]C simulations the users will have to determine on a case-by-case basis which effects they have to resolve and adjust the resolution accordingly. For example, it is perfectly possible to utilize the code without resolving the Larmor radius. However, this will cause the effect of the particles on the MHD quantities to be reduced. Since the influence of each particle is spread over multiple grid-cells, the effect of the small motions of a particle within a single cell tends to become diffusive. If the Larmor radius is not resolved, the gyrating motion of the particles around the magnetic field lines can be effectively lost. 
In order to avoid this issue we have added a refinement criterion over particle Larmor radius to the mesh refinement conditions. If the average Larmor radius of the particles within a grid falls below a pre-set fraction of the cell-size, the grid is refined.

In an explicit MHD code, timestep size and resolution are linked through the CFL condition:
\begin{equation}
\Delta t \leq ~C\,\rm{min}\biggl(\frac{\Delta~x}{v_{\rm max}}\biggr)
\end{equation}
with $\Delta~t$ the time step, $C$ a constant (usually less than one), $\Delta~x$ the size of a gridcell and $v_{\rm max}$ the maximum speed at which a signal travels through the gas. The latter is calculated as the sum of the bulk motion of the thermal fluid and wave velocities, such as sound and Alfv{\'e}n speed. 
A similar condition applies to the PIC part of the code, but with $v_{\rm max}$ being the maximum speed of an individual particle. 

Since particles will typically travel much faster than the thermal plasma, it is usually the PIC part of the code that limits the amplitude of the time step. In order to increase computational efficiency, we allow for the possibility to let the code take multiple PIC steps within a single MHD timestep. When this is done, the magnetic and electric fields are kept frozen for the duration of several PIC timesteps, until the code has reached the next MHD timestep. Obviously, this method must be used with care, but it decreases the computation time and can be applied safely for situations where the gas only responds slowly to non-thermal particles. 

An additional check is placed on the timestep. This becomes important if particles have a gyro-radius that is smaller than the size of a single grid cell, which is possible once the grid has reached the maximum level of refinement. At this stage the CFL condition no longer guarantees that the motion of the particle is properly resolved. Therefore, we also limit the timestep to a certain fraction of the gyro-time.

\section{Particle acceleration at non-relativistic shocks: parallel magnetic field case}
\label{sec-parallel}
In order to test our newly developed AMR-PI[MHD]C code, we first consider a new set-up to study the interaction between a single supra-thermal particles species (protons) and a plane shock with the magnetic field perfectly aligned with the direction of propagation of the shock. A similar simulation has been presented in \citet{Baietal:2015} where the authors have explored the impact of particle injection upon magnetic field amplification and particle acceleration at collisionless astrophysical shocks. If the physical conditions are similar to the simulation presented in \citet{Baietal:2015}, our simulation is performed in the shock frame rather than in the upstream fluid frame. Such a choice enables us to follow the evolution of the shock more easily.

\subsection{Simulation set-up}
The simulation is run in a box with a size of 240$\times$30  times a reference Larmor radius $r_{\rm g}=v_{\rm inj}/\omega_{\rm c}$, corresponding to the gyroradius of particles injected with a velocity $v_{\rm inj}$ propagating in the upstream magnetic field with a ion-cyclotron pulsation $\omega_c=q/mc B_0$ 
The base level of the grid contains $240\times 30$ grid cells. We allow the adaptive process to use three mesh levels of refinement depending on the MHD physical conditions throughout the domain. Additionally one extra level of refinement can be used in case the number of particles located in one grid approaches a maximum limit set to $7.10^4$. Enabling an extra level of refinement prevent losing particles while maintaining the computational efficiency. As a result, our grid can reach a maximum effective resolution of $3840\times 480$ cells. We chose to use a {\tt TVDLF} solver, combined with a van Leer flux limiter. Such a combination leads to a robust but precise numerical scheme able to capture the small scale features of the plasma and magnetic field.\\
\begin{figure}
\centering
\mbox{
\includegraphics[width=0.99\columnwidth]{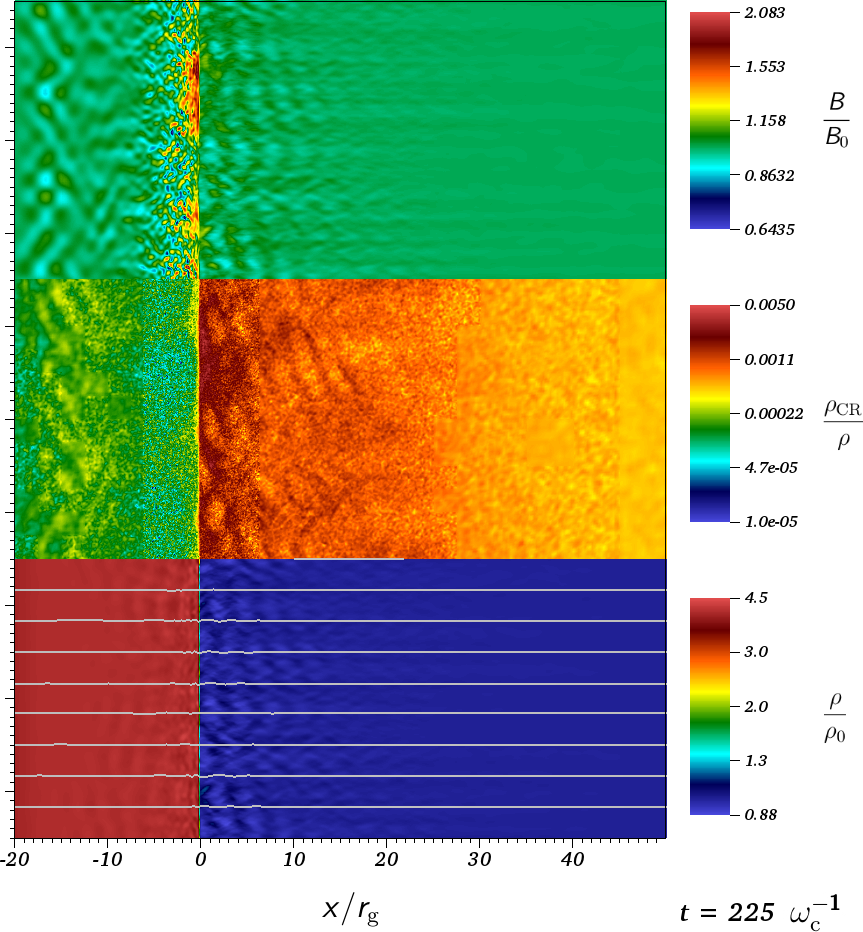}}
\caption{Color maps of the magnetic field amplitude, cosmic-ray density and plasma density in the early stages of the parallel shock simulation ($\theta_B=0$, $M_A=V_{\rm sh}/V_{\rm A0}=30$). This figure shows, from top to bottom, magnetic field strength relative to the original magnetic field, non-thermal particle charge density relative to the thermal gas density, and thermal gas mass density relative to the upstream density at the start of the simulation, combined with the magnetic field stream lines. The gas is streaming through the shock from right to left. At this point in time ($t=225\omega_c^{-1}$) the upstream medium shows the start of the streaming instability, while the downstream medium shows the onset of turbulence. Axis are normalized to $r_g$, namely the gyro radius of injected particles in the upstream medium.}
 \label{fig:parallel1}
\end{figure}
\begin{figure}
\centering
\mbox{
\includegraphics[width=\columnwidth]{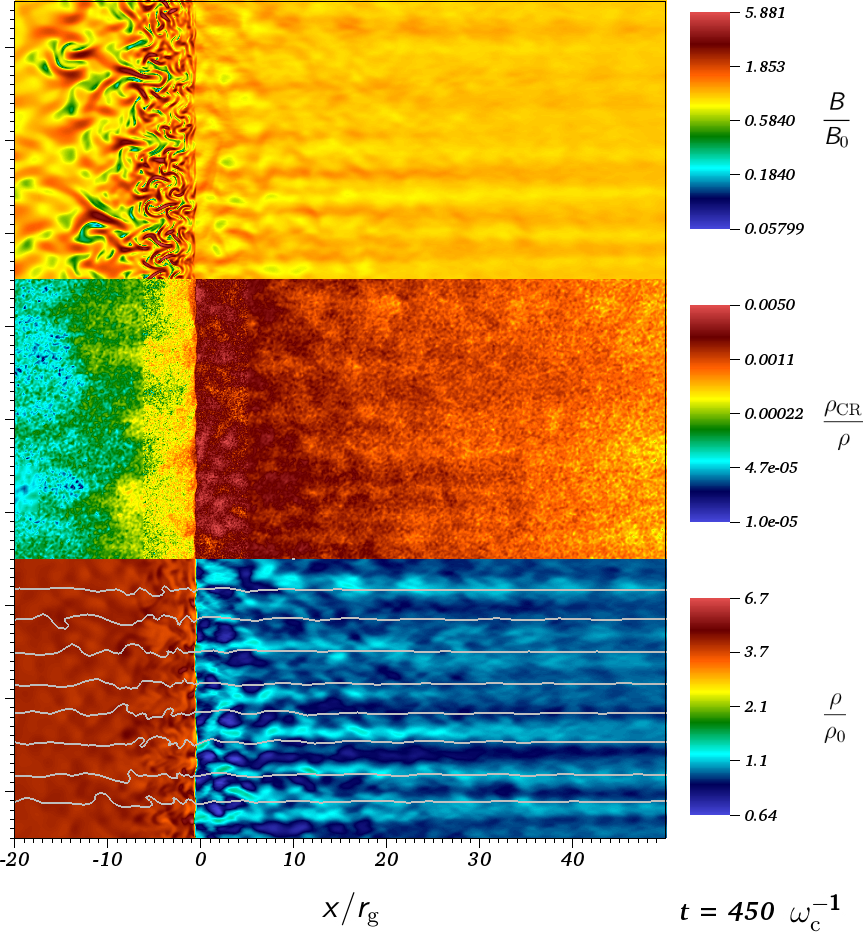}}
\caption{Similar to Fig.~\ref{fig:parallel1} this figure shows the conditions at $t=450\omega_c^{-1}$). Both the upstream streaming instability and the downstream turbulence are now fully developed.}
 \label{fig:parallel2}
\end{figure}
At the start of the simulation we set up a shock following the Rankine-Hugoniot conditions with an upstream velocity of $V_{\rm sh} = 3.10^{-3}c$ along the $x$-axis, flowing from the right $x$-boundary toward the left $x$-boundary. In order to recover an Alfv\'enic Mach number of $M_A=30$, we set the Alfv{\'e}n speed to $V_{\rm A0} = 10^{-4}c$. We also assume that in the upstream medium magnetic and thermal energies are in equipartition (i.e. $(E_B~=~E_T)$). Following \citet{Baietal:2015} we inject supra-thermal particles in the downstream medium close to the shock front with an isotropic velocity $v_{\rm inj}=3V_{\rm sh}$ in the downstream fluid frame. The number of particles injected at every timestep is determined so that the mass ratio of the injected particle to the thermal plasma mass flux is $2.10^{-3}$.

We choose inflow and outflow conditions at the right and left $x$-boundaries respectively, to accommodate the flow of the thermal plasma. For the non-thermal particles we assume that any particle that reaches the $x$-boundaries escapes from the system. For the $y$-axis, we use periodic boundary conditions for both the thermal gas and the non-thermal particles.

To ease comparison, all times are expressed in $\omega_c^{-1}$ unit. Axis length scales are measured in gyro radii units $r_g$. It is noteworthy that this fiducial length can be related to the ion plasma skin depth, namely $r_g=cv_{\rm inj}/V_{\rm A0}\omega_{\rm p,i}=90c/\omega_{\rm p,i}$ in the upstream medium. This shows that any grid belonging to the highest refinement level have MHD cells whose size is at least five times the local ion skin depth. It hence ensures that our MHD description remains valid everywhere. Finally particles are injected at a constant rate at the shock, such that $2.10^7$ particles are injected over a simulation time of $2.10^3~\omega^{-1}_{\rm c}$.

\begin{figure}
\centering
\mbox{
\includegraphics[width=\columnwidth]{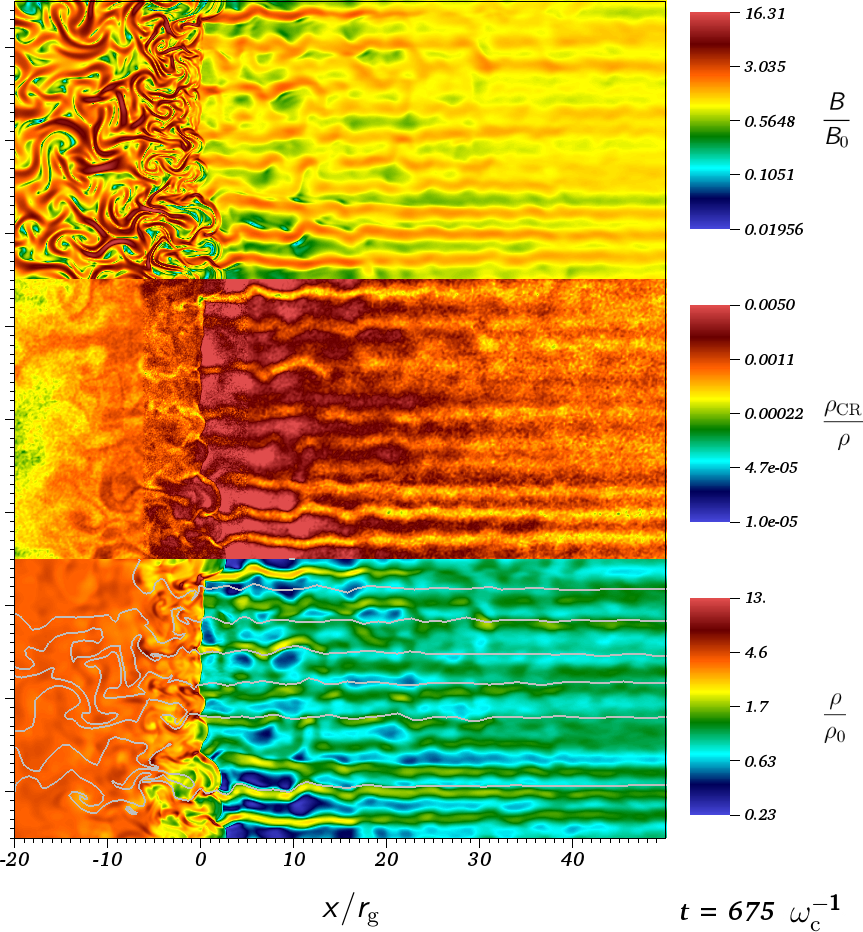}}
\caption{Similar to Figs.~\ref{fig:parallel1}-\ref{fig:parallel2} this figure shows the conditions at t=675 $\omega_c^{-1}$). The shock is warping in response to the instabilities.
}
 \label{fig:parallel3}
\end{figure}

\begin{figure*}
\centering
\mbox{
\begin{tabular}{c}
\includegraphics[width=\textwidth]{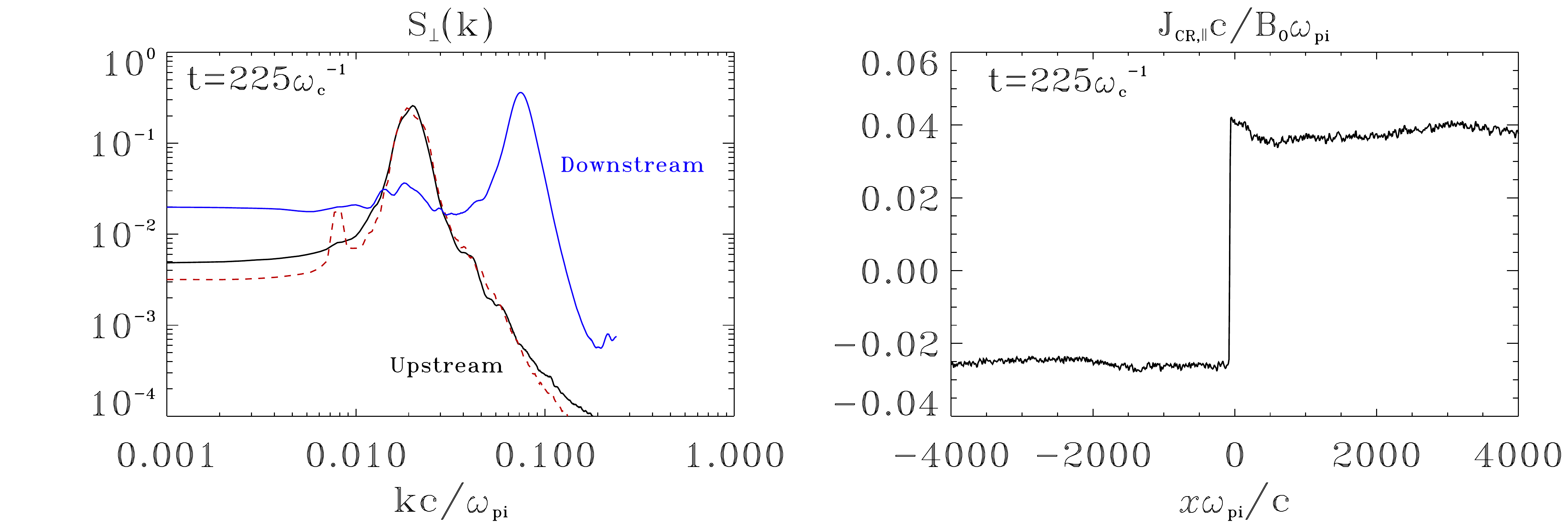}\\
\includegraphics[width=\textwidth]{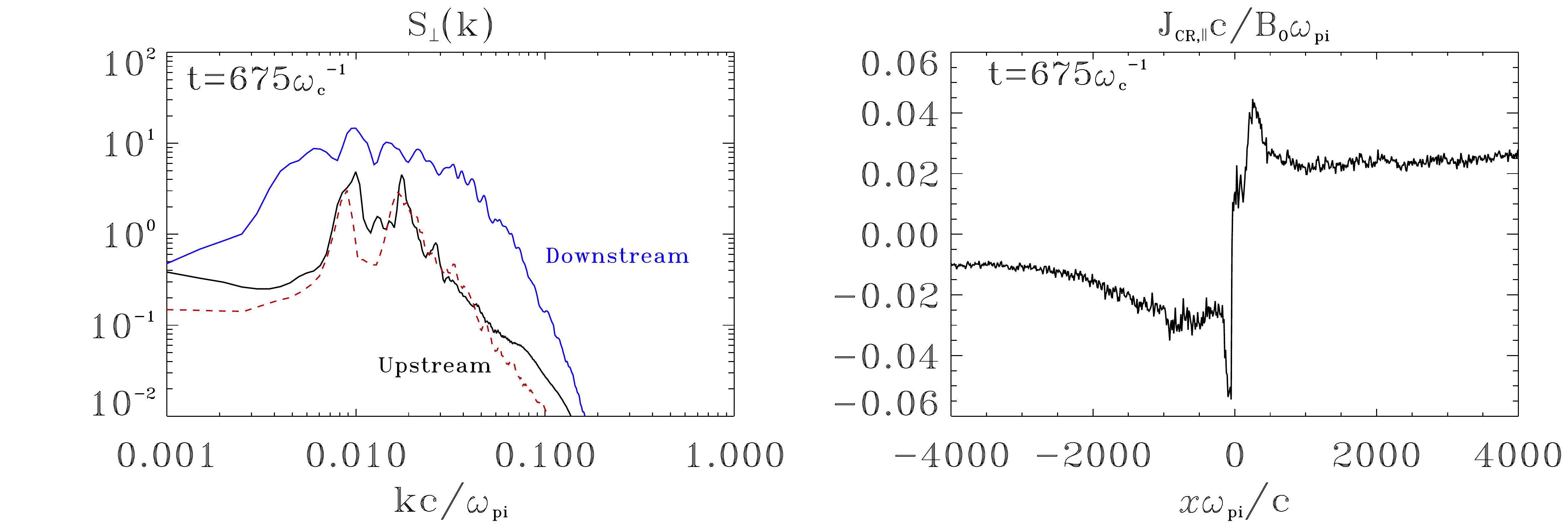}
\end{tabular}}
\caption{Transverse magnetic power spectrum and cosmic-ray current along the mean magnetic field in both upstream and downstream media. Upper panels correspond to an early stage of the simulation corresponding to Fig.\ref{fig:parallel1} while lower panels stand for a later stage corresponding to Fig.\ref{fig:parallel3}. Let us note that the red dashed line in both power spectrum plots corresponds to the upstream spectrum obtained by performing the same simulation but without any AMR refinement/coarsening, namely by setting the entire grid at the highest resolution. The relatively good agreement between upstream spectra shows that the use of AMR MHD is suitable to depict the cosmic-ray/magnetic field/thermal plasma interaction.  } 
 \label{fig:parallel3half}
\end{figure*}
\begin{figure}[t]
\centering
\mbox{
\includegraphics[width=\columnwidth]{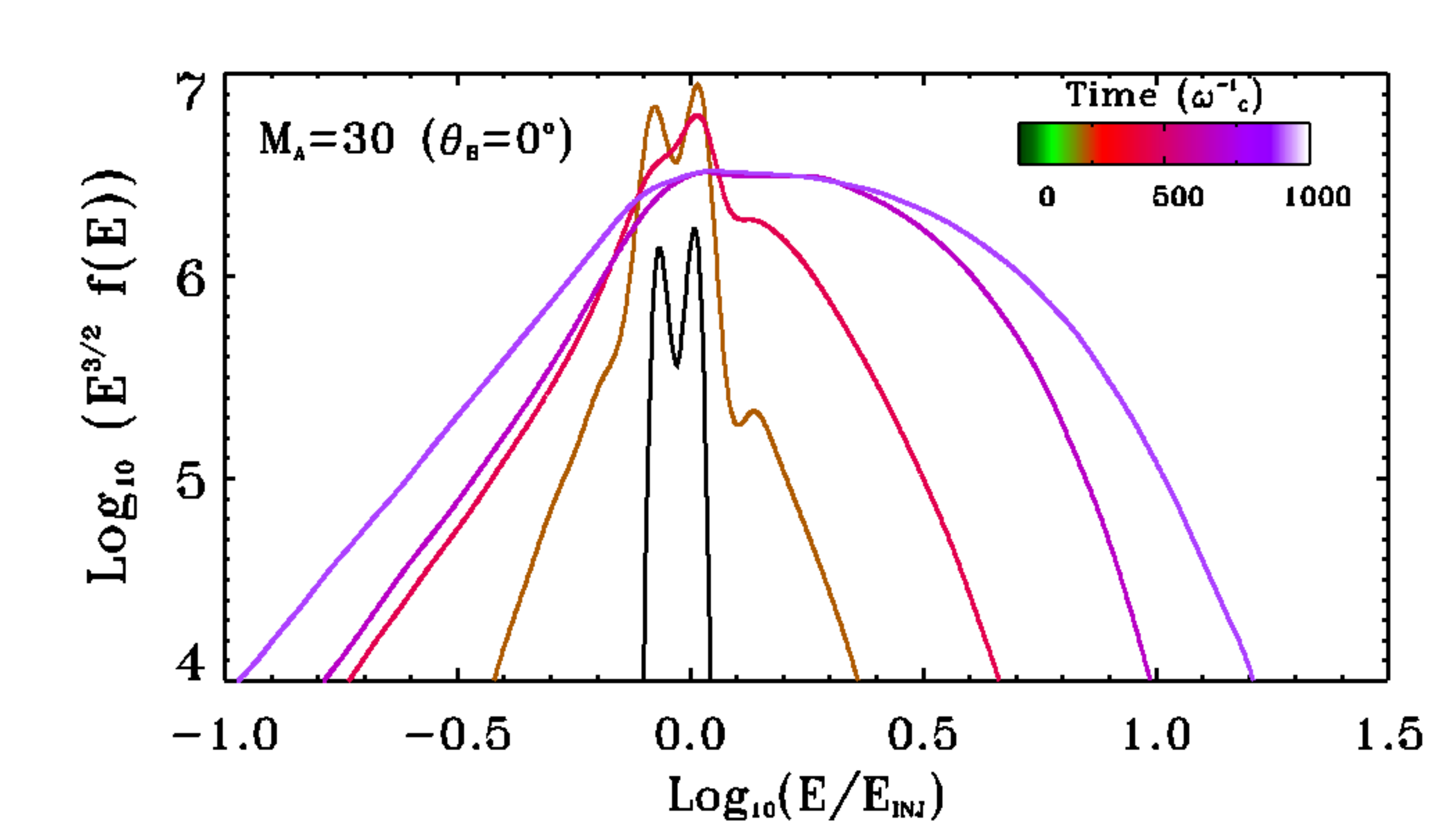}}
\caption{Energy spectra of non-thermal particles injected at energy $E_{\rm inj}/m_ic^2=4\times 10^{-5}$. The various spectra correspond to simulation displayed in Figs.~\ref{fig:parallel1}-\ref{fig:parallel2}-\ref{fig:parallel3}. A non-thermal tail is forming in the late stages of the simulation tending to a power-law spectrum in agreement with a diffusive shock acceleration process.
}
 \label{fig:spectra_para_M30_6}
\end{figure}

\begin{figure*}{t}
\centering
\mbox{
\includegraphics[width=0.33\textwidth]{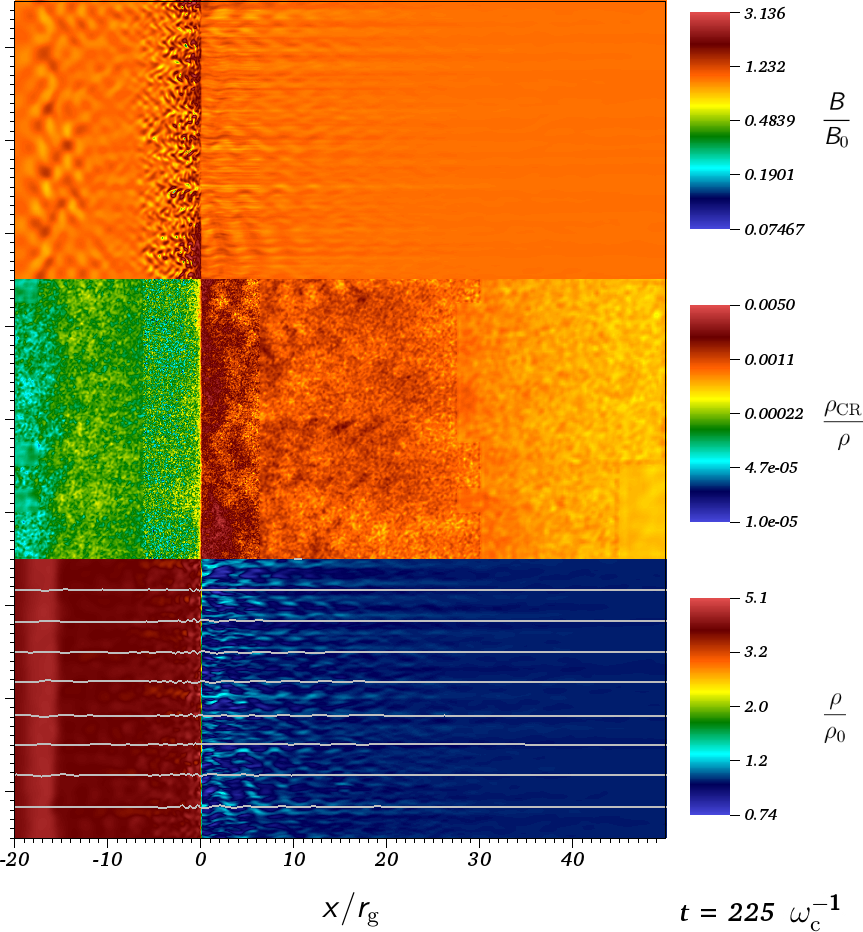}
\includegraphics[width=0.33\textwidth]{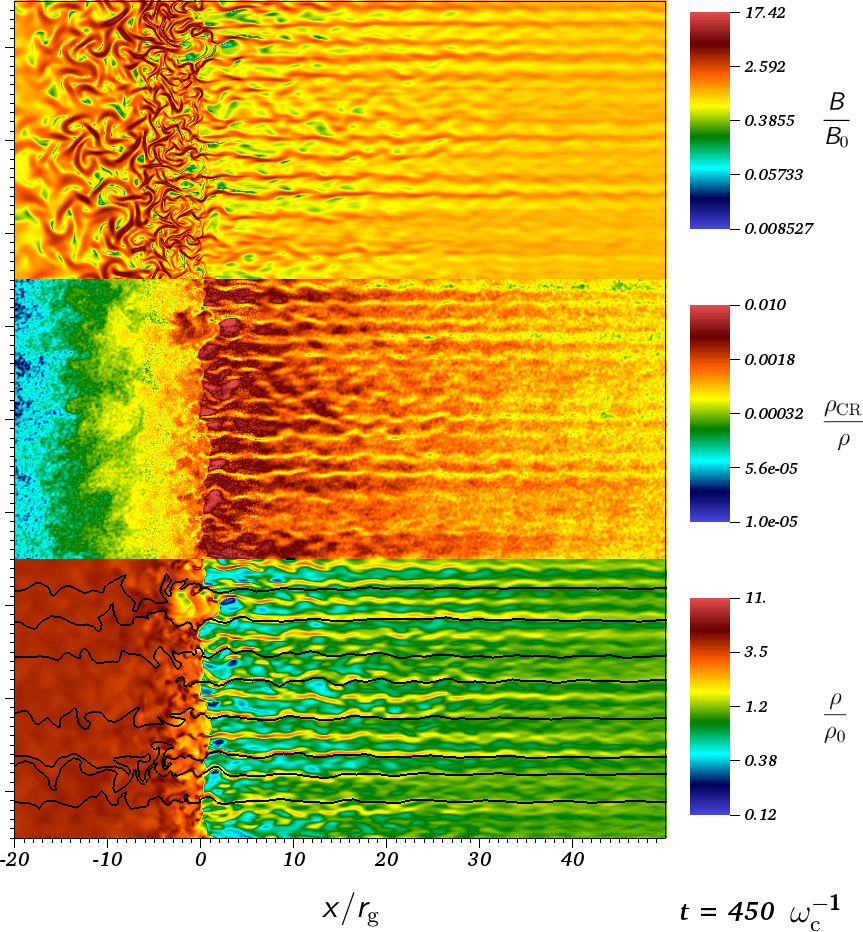}
\includegraphics[width=0.33\textwidth]{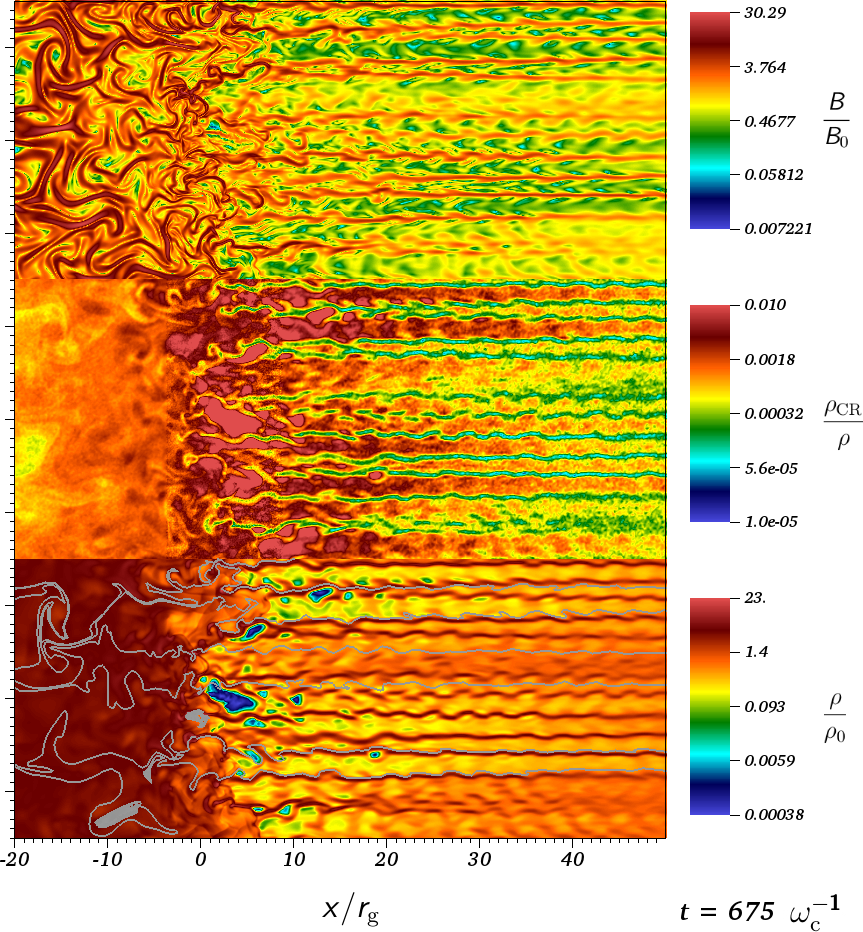}}
\caption{
Same plots than in Fig.\ref{fig:parallel1} showing the temporal evolution of a high Alfv\'enic Mach number shock ($\theta_B=0$, $M_A=300$). In all the stages of the simulation we recover the same patterns than in the slower shock simulation displayed in Figs.\ref{fig:parallel1}-\ref{fig:parallel2}-\ref{fig:parallel3}. However, one difference arises, namely the amplification of the magnetic field. In this simulation, the average magnetic field in the downstream medium has been amplified by a factor three with respect to slow shock case. This result is in relative good agreement with the pure kinetic simulations of \citet{Caprioli14a, Caprioli14b}.}
 \label{fig:parallel4}
\end{figure*}

\subsection{Results}
From the start, the effect of charged particles on the magnetic field varies considerably between upstream and downstream regions as shown in Fig.~\ref{fig:parallel1}, which shows at time $t=225\,\omega_c^{-1}$: the gas density relative to the initial upstream density $\rho_0$, the particle density relative to the gas density, and the absolute magnetic field strength relative to the initial upstream magnetic field $B_0$.

As particles move upstream, they spiral around the field lines, initiating the NRS instability \citep{Bell04}, which is characterized by a filamentary structure that runs parallel to the field lines and the flow. Because the bulk motion of the gas is super-Alfv\'enic, the motion of the thermal plasma dominates over the fluctuations in the magnetic field, keeping the instabilities aligned with the flow.
On the other hand, the effect in the downstream region is much more complex. Although the downstream flow is not sub-Alfv{\'e}nic, the difference between bulk-velocity and Alfv\'en speed is much reduced by the shock, allowing the magnetic fluctuations to start moving perpendicular to the flow. As a result, the fluctuations look random, both in strength and direction. They neither follow the field lines nor the flow. The strongest magnetic fluctuations occur downstream, just behind the shock. 
Here the effect of the upstream and downstream instabilities is combined, with the upstream instability being enhanced as it is compressed by the shock. One way to clearly identify this instability process is to compute the power spectrum of magnetic field fluctuations transverse to the mean magnetic field. Indeed, as predicted by the linear theory of the instability, the fastest growing mode of the instability is controlled by the strength of the supra-thermal particle current as $k_{\rm max}=J_{\rm CR}/2B_0$. The upper panels of Fig.\ref{fig:parallel3half} shows at $t=225~\omega_{\rm c}^{-1}$ the transverse power spectrum and the particle current. It clearly appears that upstream the power spectrum peaks at the expected wave number, namely $k_{\rm max} \simeq 0.02 \omega_{\rm pi}/c$. The transverse power spectrum is defines as
\begin{equation}
S_{\perp}(k)=\frac{|\delta \hat{B}_{\perp}^2(k)|}{B_0^2}
\end{equation}
where $k$ is the wave-vector oriented along the mean magnetic field $B_0$ and $\delta\hat{B}_{\perp}(k)$ stands for the Fourier transform of the transverse magnetic field.
The downstream counterpart of the spectrum exhibits the same behavior but with a dominant mode roughly four times larger that for the upstream spectrum. This behavior is consistent with Alfv\'en waves wavevector transmission at a parallel non-relativistic shock \citep{Vainio98}. On Fig.\ref{fig:parallel3half}, the magnetic power associated with transverse magnetic fluctuations remains small compared to the mean magnetic field energy density, a clear sign that the instability is still in a linear growth phase. 

As a result of these two different patterns, the motion of the particles changes as well. Upstream the particles can effectively escape from the domain by moving upstream, along the field lines, until they hit the boundary. However, because the downstream magnetic field acquires significant $y$- and $z$-components, the particles that move downstream are delayed as their motion is twisted to follow the local magnetic field, hence experiencing a random walk. 
As a consequence, the turbulent zone effectively becomes a barrier that few particles are able to pass. Instead, they are either reflected into the upstream medium, or become trapped inside the turbulent zone downstream. 
Conversely, owing to the parallel filamentary nature of the instabilities upstream, the particles that move upstream do not encounter an effective magnetic barrier and can escape quite easily. However, we should note that the particle confinement is influenced by the 2-D nature of our simulations as the non-linear growth of the instability may differ from  full 3D computations (see \citet{Belletal:2013} where authors considered both 2D and 3D Vlasov-MHD simulations).

When the instabilities increase in strength, the thermal plasma starts to respond, following the local motion of the magnetic field. Upstream, this leads to an enhanced filamentary structure, whereas downstream the medium becomes increasingly turbulent. This turbulent zone effectively traps the non-thermal particles, increasing their local density, which in turn increases the instability (Fig.~\ref{fig:parallel2}.)
Meanwhile, the amplification of the magnetic field rises until, locally, it reaches values of approximately fifteen times the original strength. It is noteworthy that the transverse magnetic power spectrum in the late phase of the simulation still exhibits dominant modes in agreement with the NRS instability in the precursor while downstream fluctuations have increased well beyond the initial magnetic field strength and are spread over a large range of wavevectors. This last feature is of great interest for particle acceleration as it is mandatory to create large scale magnetic fluctuations in order to induce a random motion of more and more energetic supra-thermal particles \citep{Zirakashvili08}. 

At later times, the particle current starts to influence the shock itself, which starts to move as it adjusts to local variations in density, velocity and pressure. Eventually, the shock becomes warped to the point that it becomes impossible to define its exact location (Fig.~\ref{fig:parallel3}.) At this moment we stop the simulation because we are not able to inject particles at the shock front anymore because the injection process becomes dependent of the shock structure itself. Longer term simulations require a more careful analysis of the injection procedure and are scheduled for a future work.

Fig.(\ref{fig:spectra_para_M30_6}) shows the energy spectra of supra-thermal particles contained within the computational domain at different stages of the simulation. In the early stages, the particle energy distribution remains peaked around the injection energy while as the simulation progresses particle distribution broadens. Beyond $t=350~\omega_c^{-1}$ a high-energy tail appears and tends to form a power-law whose index is in agreement with diffusive shock acceleration theory, namely $f(E)\propto E^{-3/2}$ where $E$ is the kinetic energy of the particles. In our simulation, the tail formation stalls beyond $t=600\,\omega_c^{-1}$ mainly due to the deformation of the shock front leading to a deficient particle injection.  

Finally we have performed a simulation without any AMR refinement while setting all the computational domain to the most refined level. With such settings, we have basically recovered all the results from the previous simulation. On Fig.(\ref{fig:parallel3half}) we plot in dashed-line the power spectrum of the transverse magnetic fluctuations obtained in the case where no AMR grid is triggered: the downstream magnetic spectrum is recovered and the same dominant mode of the turbulence is present in each phase of the simulation. Such a test proves that the use of an AMR grid is suitable to perform PI[MHD]C simulations and even recommended for future applications requiring a much larger computational domain.

\subsection{Higher Mach number simulations}
We repeat above simulations for an increased Alfv{\'e}nic Mach number ($M_A=300$), which corresponds more closely to the type of shock expected of the early phase of a supernova remnant expansion into the ISM. 
Higher Mach numbers are achieved by increasing the velocity of the thermal fluid by an order of magnitude. The initial velocity of newly injected particles is increased by the same amount. 
In order to keep the other quantities (density, magnetic field strength) identical, we also increase the Larmor radius by a factor ten and increase the size of the simulation box accordingly, while maintaining the PIC resolution (the number of grid cells per Larmor radius).
\begin{figure}
\centering
\mbox{
\includegraphics[width=\columnwidth]{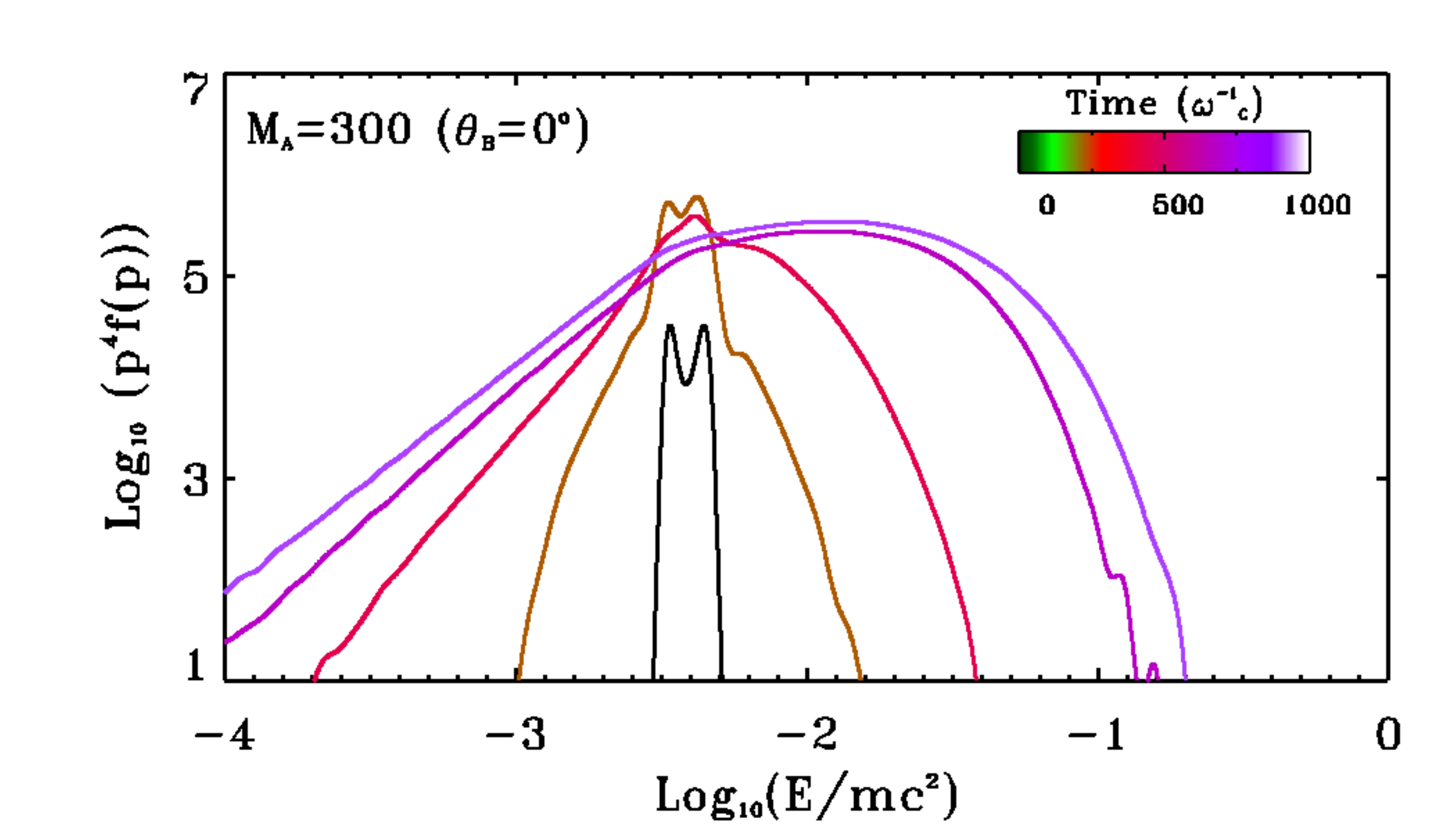}}
\caption{Energy spectra of non-thermal particles injected at energy $E_{\rm inj}/mc^2=4\times 10^{-5}$. The various spectra correspond to simulation displayed in Fig.~\ref{fig:parallel4}.
}
 \label{fig:spectra_para_M300_6}
\end{figure}
As shown in Fig.~\ref{fig:parallel4}, the behavior of the shock follows the same pattern as for the $M_A=30$ case, albeit that the local magnetic field amplification is approximately three times as strong. This last result is in agreement with the pure kinetic (hybrid PIC) computations presented by \citet{Caprioli14b} where the authors have investigated the role of the Alfv\'enic Mach number of the shock over the magnetic field amplification near a non-relativistic shock. We have considered here a $M_A=300$ shock and found that the amplified magnetic field at saturation follows the same relation where $<B_{\rm tot}^2/B_0^2>\sim 0.45 M_A$. Such an agreement between full PIC simulation and PI[MHD]C computations highlights the viability of the latter method. It is important to mention that we have performed simulations of shocks exhibiting an Alfv\'enic Mach number up to $M_A=3000$ thanks to the robustness of the PI[MHD]C method and again recover the basic same patterns than in the $M_A=30$ case. In a 
forthcoming study we will explore the process of magnetic field amplification and particle acceleration in relativistic shocks.

\section{Particle acceleration at magnetically oblique shocks}
\label{sec-Bfieldangle}
\begin{figure*}
\centering
\mbox{
\includegraphics[width=1.99\columnwidth]{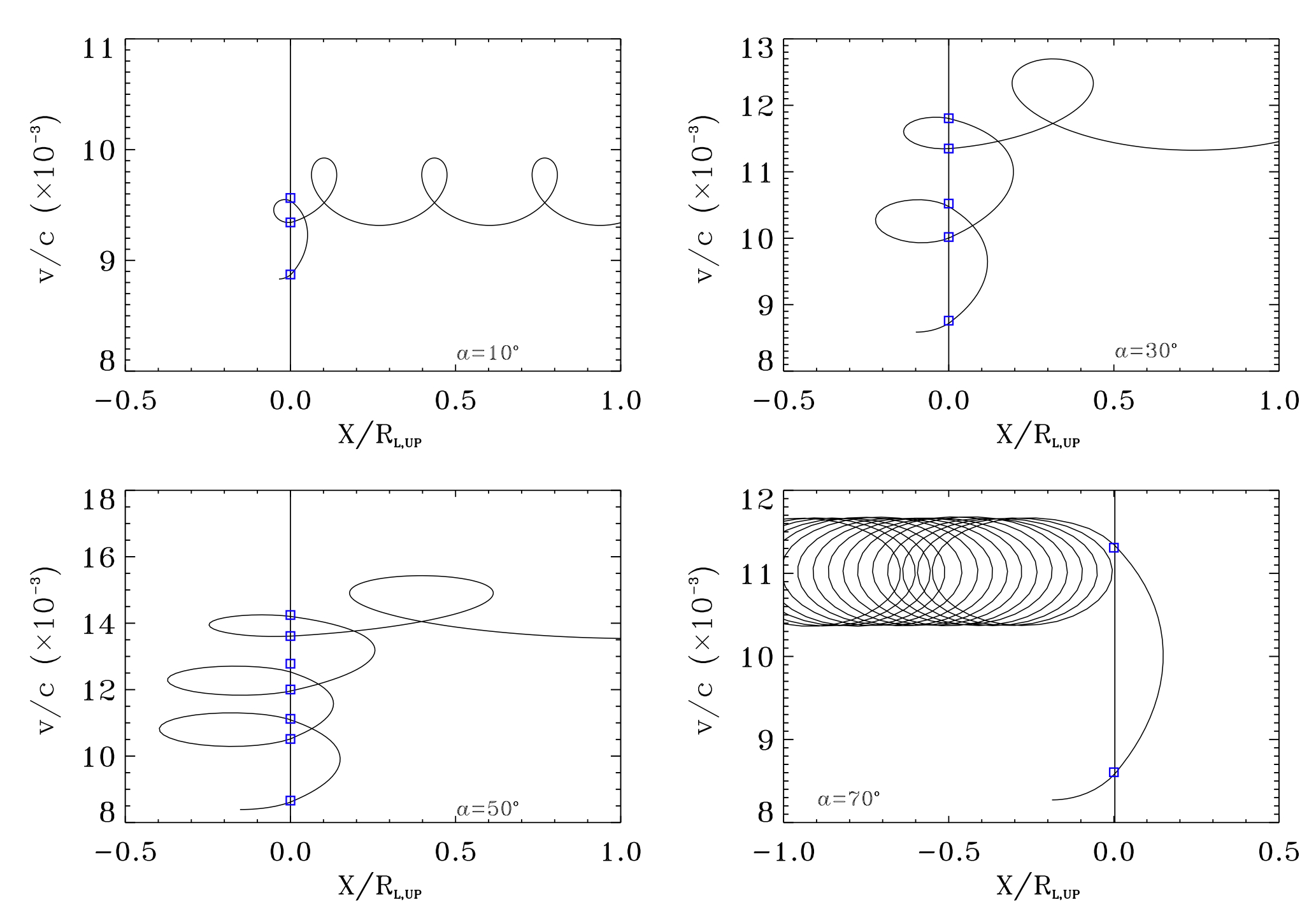}}
\caption{Evolution of the velocity of four supra-thermal particles injected in the downstream medium with an initial velocity of $3V_{\rm sh}$ in the downstream frame. Each particle is injected with a different pitch angle $\alpha$ with respect to the oblique magnetic field $(\theta_B=70^o$). Blue rectangles stand as energy gains predicted by the scattering of particle upon a thin shock when considering initial properties of the particles.
}
 \label{fig:accel}
\end{figure*}
In this section we consider the process of particle acceleration near oblique astrophysical shocks where a mean magnetic field is inclined at an angle $\theta_B$ with the propagation direction of the shock. Pure kinetic computations have considered such a configuration and have concluded that when $\theta_B>45^o$, no magnetic field amplification nor significant particle acceleration is taking place (see for instance \citet{Caprioli14a} and reference therein). The goal of the simulations presented in this section is to put this statement to the test by investigating over a long period of time the interplay between supra-thermal particles and an oblique super-Alfv\'enic MHD shock. To do so, we choose to consider a $\theta_B=70^o$ oblique shock with various Alfv\'enic Mach numbers fulfilling the Rankine-Hugoniot jump conditions in the shock frame. 

\subsection{Moderate Alfv\'enic Mach case ($M_A=30$)}
The injection process of supra-thermal particles follows the same recipe than in the parallel case, namely injecting particles with an isotropic velocity $v_{\rm inj}=3V_{\rm Sh}$ in the downstream frame. Injection occurs close to the shock front in the downstream medium. As mentioned in previous studies (see e.g. \citet{Caprioli15} and references therein), injected particles reaching the shock can be reflected at the shock front because of the obliquity of the magnetic field. Return to the shock is then quite unlikely in the early phase of the magnetic field amplification as there are no significant magnetic fluctuations that can alter the trajectory of the particles. The various toy models presented in previous studies agree to state that shocks exhibiting oblique magnetic field verifying $\theta_B>45^o$ are not allowing a sufficient amount of particle to get into the upstream medium to trigger NRS instability. 

\begin{figure*}
\centering
\mbox{
\includegraphics[width=0.32\textwidth]{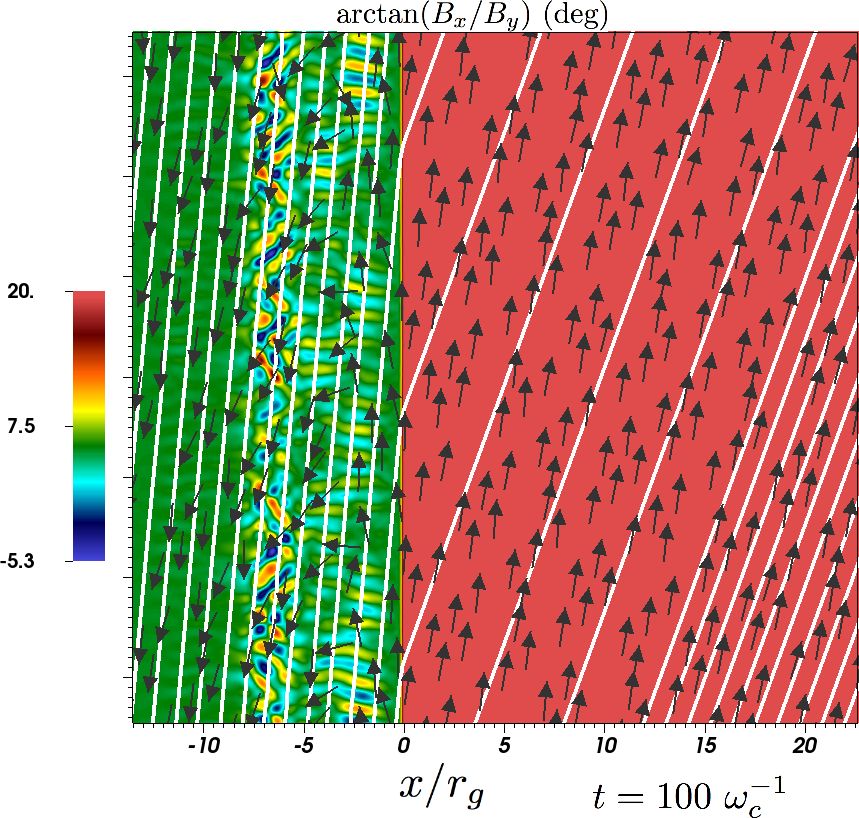}
\includegraphics[width=0.32\textwidth]{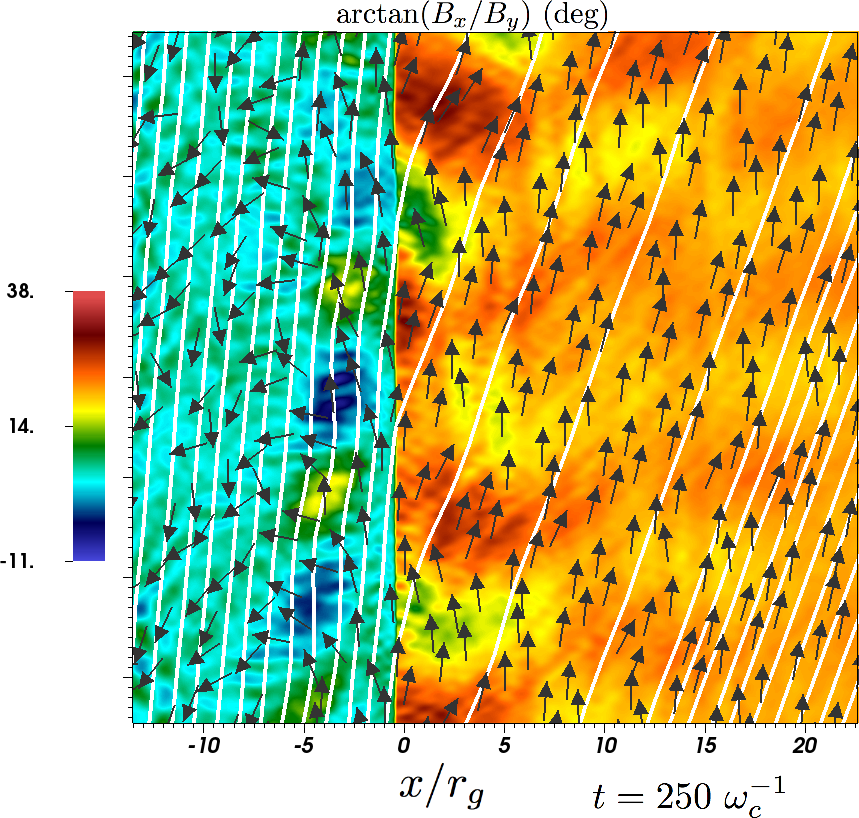}
\includegraphics[width=0.32\textwidth]{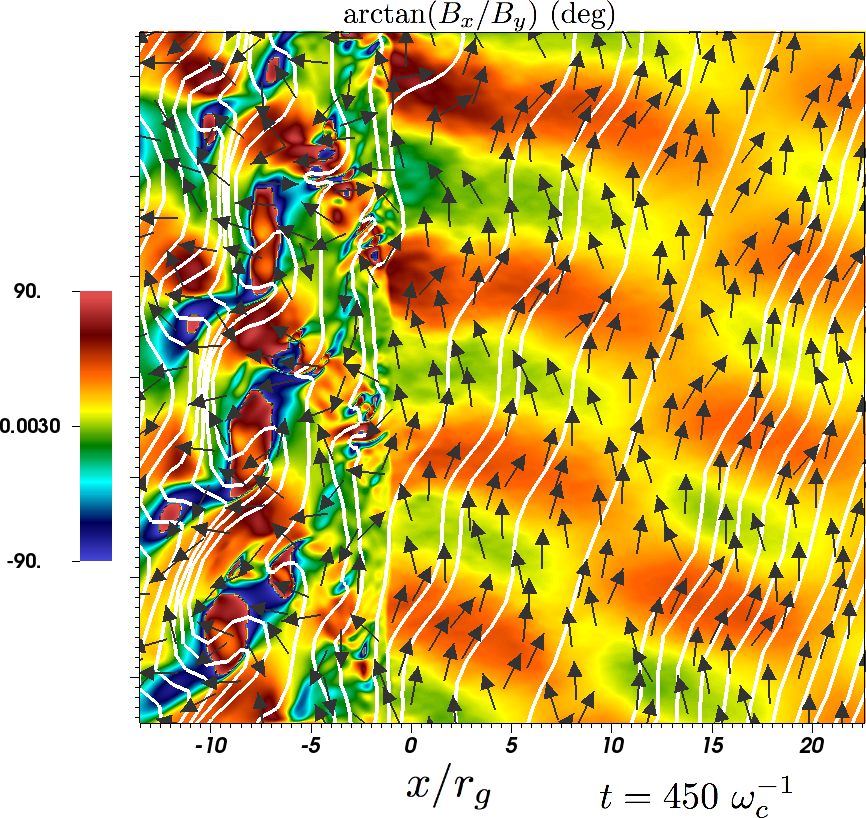}}
\caption{Temporal evolution of the magnetic field lines for the $M_A=30$ oblique shock. The colormap represents the opening angle of the magnetic field lines in the $x-y$ plane while white lines stand as the magnetic field lines. Black arrows indicate the orientation of the cosmic-ray current in the same plane. In such oblique configuration, the cosmic-ray current is not parallel to the magnetic field as a drift motion is enforced by the electromotive field. The initial shock drift acceleration enables accelerated particles to flow in the upstream medium but does not trigger the NRS instability as the cosmic-ray electrical current is slowly propagating in the upstream medium compared to a parallel shock configuration}. The NRS instability eventually occurs in the downstream medium where a large wavelength magnetic perturbation leads to the corrugation of the shock hence triggering particle acceleration.
 \label{fig:Current_perp}
\end{figure*}
\begin{figure*}
\centering
\mbox{
\includegraphics[width=\columnwidth]{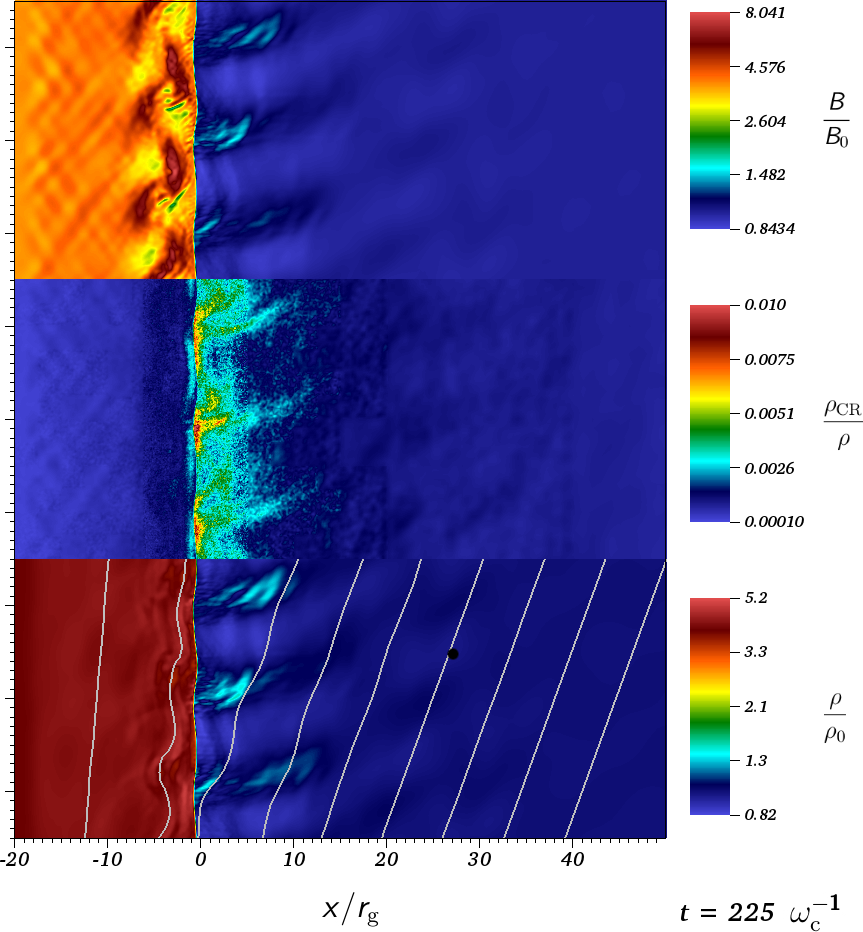}
\includegraphics[width=\columnwidth]{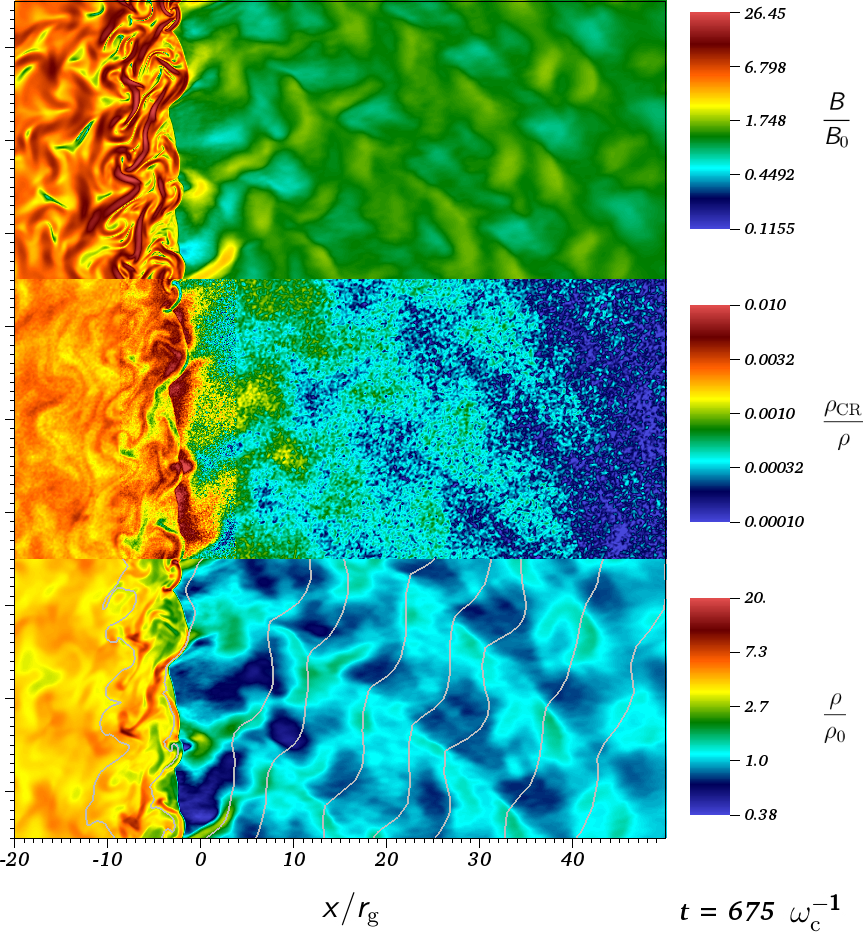}}
\caption{Similar to Figs.~\ref{fig:parallel1} and \ref{fig:parallel4}, but with $M_A=30$ and the angle between the magnetic field and the direction of the flow $\theta_B=70^{\rm o}$. These two snapshots exhibit the late stages of the oblique shock simulation where the corrugation of the shock front, induced by a NRS instability behind the shock, leads to particle acceleration as part of the shock display physical properties consistent with parallel-type shock configuration.
}
 \label{fig:20deg1}
\end{figure*}
\subsubsection{Initial shock drift acceleration}
As we inject particles near the shock, we notice that in the very first stages of the simulation, the particle energy distribution function, originally peaked around injection energy, is now expanding rapidly up to a point where the most energetic particles reaches ten times the injection energy (see Fig.\ref{fig:spectra_perp_M30}). This distribution function then stalls until approximately time $t=200\omega_c^{-1}$. We checked the path of many particles and found that the accelerated particle have experienced multiple shock encounters while many particles have roughly kept their initial energy.     

On Fig.(\ref{fig:accel}), we have displayed the velocity variation as a function of the $x$ coordinate of four particles, each one of them being injected with a different pitch-angle $\alpha$ with respect the downstream magnetic field. Some of these particle experienced multiple shock crossings leading to a so-called shock drift acceleration process (SDA, see e.g. \citet{Decker:1988}). Since numerical MHD schemes display shock discontinuities as a velocity transition occurring over a few MHD cells, one may question the ability of PI[MHD]C code to properly depict the particle acceleration process. 
In order to test the accuracy of the code, we have added on top of the particle  trajectory in the $x-v$ space in Fig.\ref{fig:accel} the velocity jump gained by each particle going from downstream to upstream and then back to downstream. This estimate is straightforward as we record the velocity and pitch-angle of the particle when crossing the shock and compute the velocity jump accordingly to the conservation of the particle velocity in the upstream frame, namely
\begin{eqnarray}
\Delta v &=& ((v_o'\mu_++U_u)^2+v_o'^2(1-\mu_+^2))^{1/2} \nonumber \\
&-& ((v_o'\mu_-+U_u)^2+v_o'^2(1-\mu_-^2))^{1/2}
\end{eqnarray}
where $v_o'$ is the velocity of the particle when entering the upstream medium measured in the upstream frame while $\mu_-$ and $\mu_+$ are the cosine of the pitch-angle of the particle when entering the upstream medium and going back to the downstream medium respectively, $U_u$ is the flow speed in the shock restframe. The good agreement between the few $x-v$ paths and the aforementioned prediction shows that the numerical MHD shock transition is narrow enough to  accurately describe the particle acceleration via SDA. This is obviously achieved thanks to the numerical setup of the simulation leading to the Larmor radius of the injected particle being larger than the MHD shock thickness.
\begin{figure}
\centering
\mbox{
\includegraphics[width=\columnwidth]{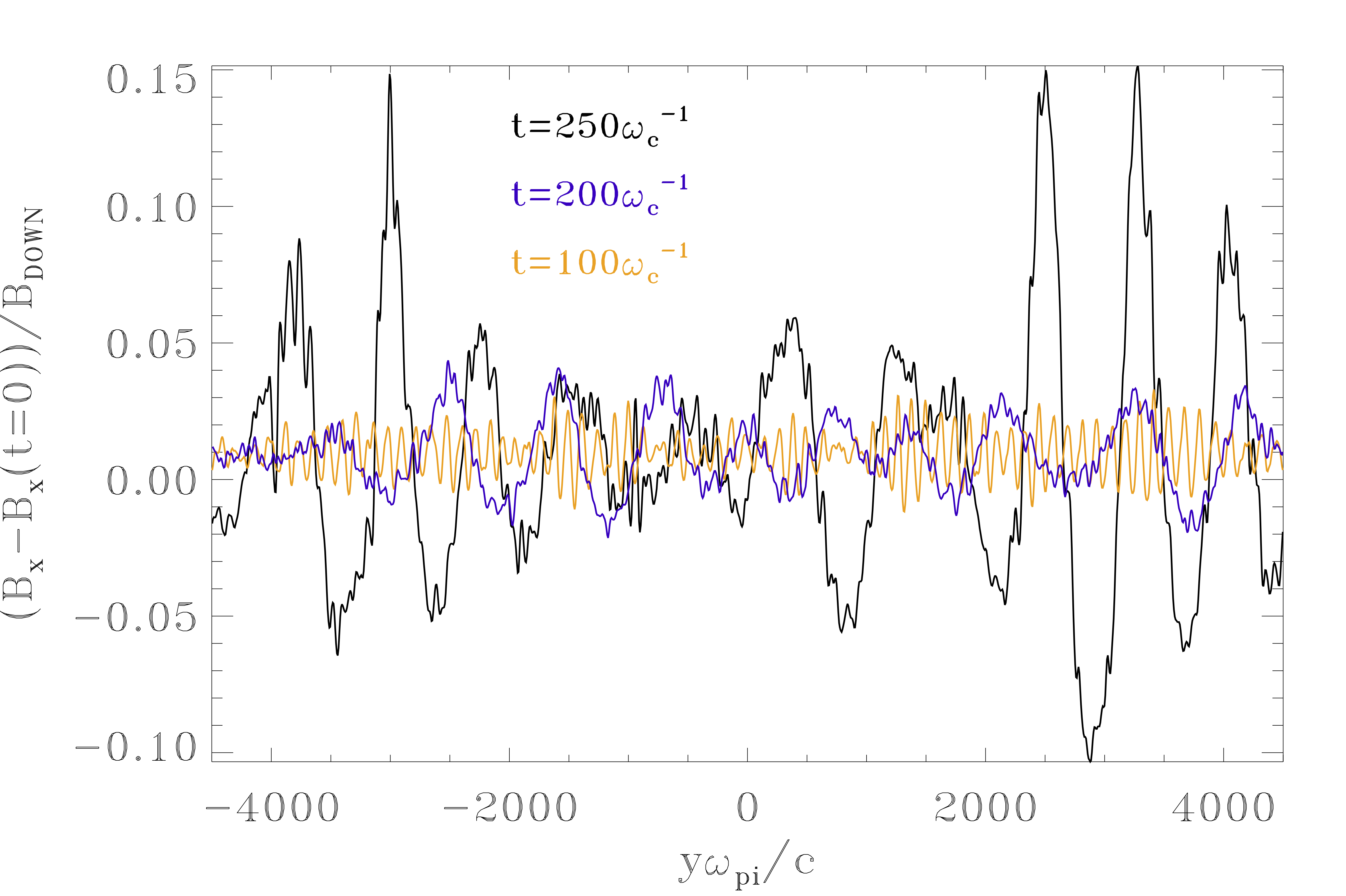}}
\caption{Variation of $\delta B_x$ along the $y$-axis just behind the shock in the downstream region. Initially, small wavelength perturbations are present due to the SDA process at work at the shock. Later on, a large scale oscillation grows and dominates the entire magnetic structure leading to the corrugation of the shock. This large-scale modulation of the magnetic field is induced by a NRS instability triggered by the weak supra-thermal electric current.
}
 \label{fig:profile_B_perp}
\end{figure}

As soon as we inject supra-thermal particle behind the shock, they experienced SDA leading to a shift from the initial mono-energetic energy distribution to a wider energy distribution as one can see on Fig.\ref{fig:spectra_perp_M30}. We do observe that particles either gain or lose energy depending on their injection orientation and that the highest velocity obtained through SDA is roughly three times the injection velocity, namely $9V_{\rm sh}$. Such process has already been observed in hybrid simulations (see e.g \citep{Caprioli14a}) but the use of PI[MHD]C techniques provides here an advantage as the pool of supra-thermal particles in our simulations is far larger than in hybrid computations. Indeed, we do inject up to 20 millions supra-thermal particles in the span of the simulation. In order to reach such statistic, an hybrid simulation would have to consider roughly 20 billions particles crossing the shock. In our simulation, the most energetic particles are able to flow into the upstream medium as 
they are able to counteract the drift velocity induced by the presence of the electromotive field. As a result of the drift motion, the cosmic ray current is no longer parallel to the magnetic field but almost vertical leading to a slowly expending cosmic ray charge distribution in the upstream medium.

\subsubsection{Corrugation of the shock front}
\begin{figure*}
\centering
\begin{tabular}{c}
\includegraphics[width=0.8\textwidth]{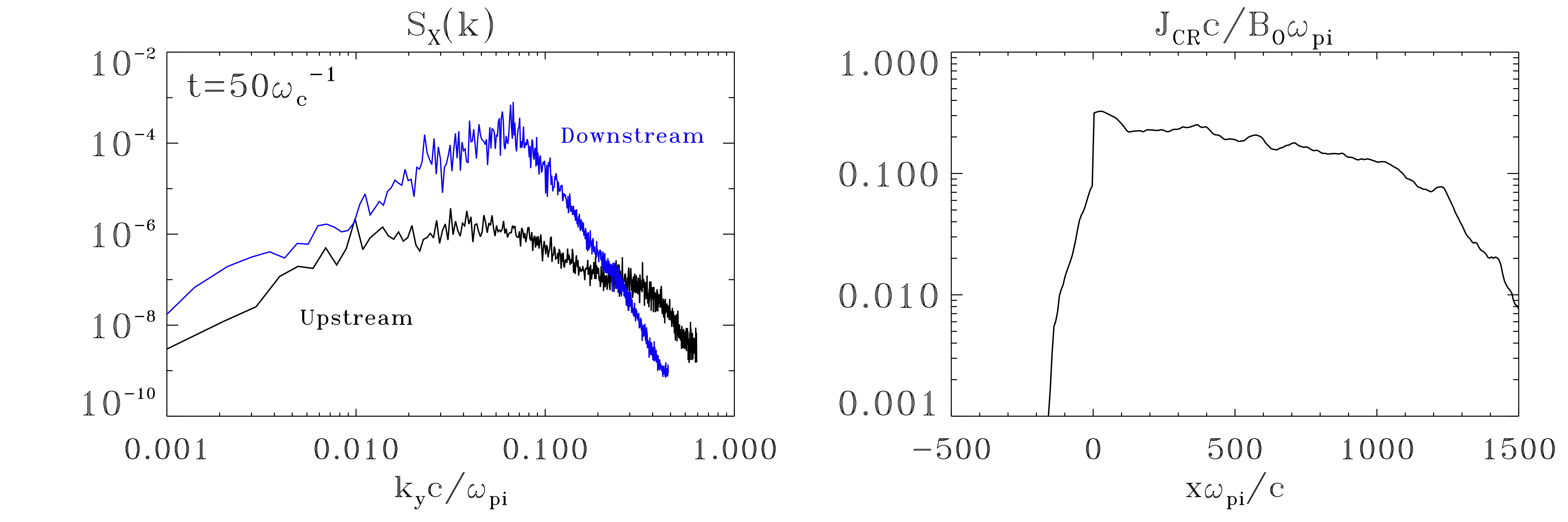}\\
\includegraphics[width=0.8\textwidth]{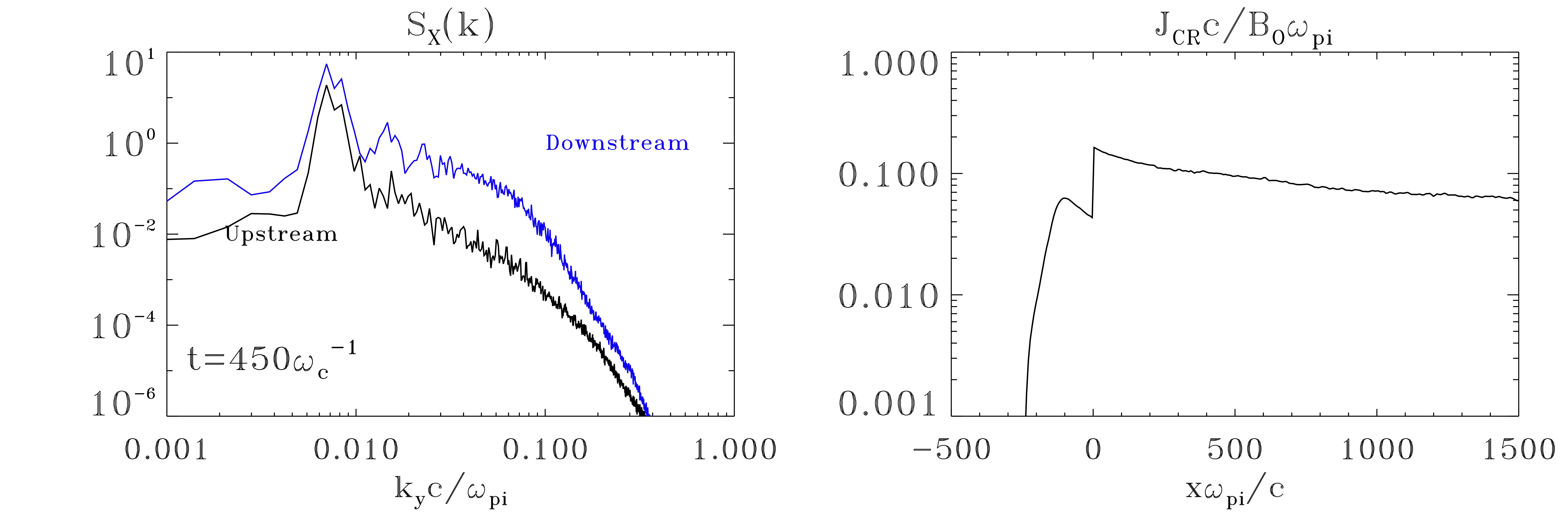}
\end{tabular}
\caption{Upstream and downstream magnetic power spectrum of the magnetic field component parallel to the flow in (left) a early phase and in a (right) later stage of the simulation. In the early phase the upstream medium exhibits no dominant mode showing that the NRS instability is not triggered despite the presence of an upstream current. We do see however fluctuations in the downstream medium related to the injected particle being process through SDA as fluctuations occur just behind the shock front. In the later phase of the simulation a large wavelength mode is visible in  both the downstream and upstream magnetic spectra. Such large wavelength oscillation do correspond to the corrugation wavelength of the shock and is induced by a NRS instability occurring in the downstream medium. The small current present just behind the shock are shown on the right column.}
 \label{fig:Spectrum_perp}
\end{figure*}
The propagation of supra-thermal particles in the upstream medium leads to a strong filamentary instability when the magnetic field is nearly parallel to the shock front normal. During the initial expansion of supra-thermal particles in the precursor of the oblique shock, we do not observe any NRS instability features, apart from very small amplitude fluctuations in the magnetic field. \citet{Bell05} have presented a derivation of the dispersion relation of the instability for any obliquity of the magnetic field and stated that this instability can occur provided that the magnetic field is not perpendicular to the incoming flow. \citet{Reville12} completed the description of the instability by taking into account the back-reaction of the plasma upon the supra-thermal particles in a parallel shock framework. In such context, one can summarize the whole instability process by considering the link between the supra-thermal current and the potential vector component parallel to the flow, namely 
$J_{\rm part}\propto A_{\parallel}$. If a local maximum of $A_{\parallel}$ is met at some location, the magnetic force induced by supra thermal particles is pushing out the plasma material from the initial location hence expending the region where $A_{\parallel}$ is maximal. The related increase of the supra-thermal current will then lead to a runaway process triggering the filamentation instability.  

In an oblique shock context, the local physical conditions are quite different. Indeed obliquity angles $\theta_B$ larger than $45^o$ leads to the aforementioned shock drift acceleration process when injecting particles near the shock front. Particles crossing the shock front to propagate in the precursor are selected through the SDA process. In order to counterbalance the drift motion induced by the local electromotive field, particles must have a sufficiently high velocity module and pitch-angle. Such selection leads to a slowly expanding cosmic-ray electrical current in the upstream medium. During the beginning of the simulation, this upstream electrical current does not reach a steady state conversely to the parallel shock case. This configuration differs from the framework  of the linear analysis of \citet{Bell05} where  a constant cosmic-ray current over space and time is assumed. The varying current is likely to prevent a dominant wavelength to arise in the upstream medium so that no filamentary 
structure appears (see Fig.\ref{fig:Spectrum_perp} upper left). In the meantime, downstream small amplitude magnetic variations can be seen in the close vicinity of the shock (Fig.\ref{fig:Current_perp}). These fluctuations are induced by the streaming of the particle experiencing the SDA process. 

In the later phase of the simulation, a large scale oscillation of the magnetic field arises just behind the shock in the downstream medium (see middle panel of Fig.\ref{fig:Current_perp}). We display on Fig.\ref{fig:profile_B_perp} the average variation of $B_x$ just behind the shock at three stages of the simulation. One can clearly see that as time goes by, a long wavelength oscillation occurs in the magnetic field (the same profile also applies to $B_z$). Such an oscillation unbalances the equilibrium of the shock front leading to a corrugation of the shock whose wavelength matches the one from the magnetic oscillation (see Fig.\ref{fig:20deg1}).   
\subsubsection{Particle acceleration near the corrugated shock}
 
\begin{figure}
\centering
\mbox{
\includegraphics[width=\columnwidth]{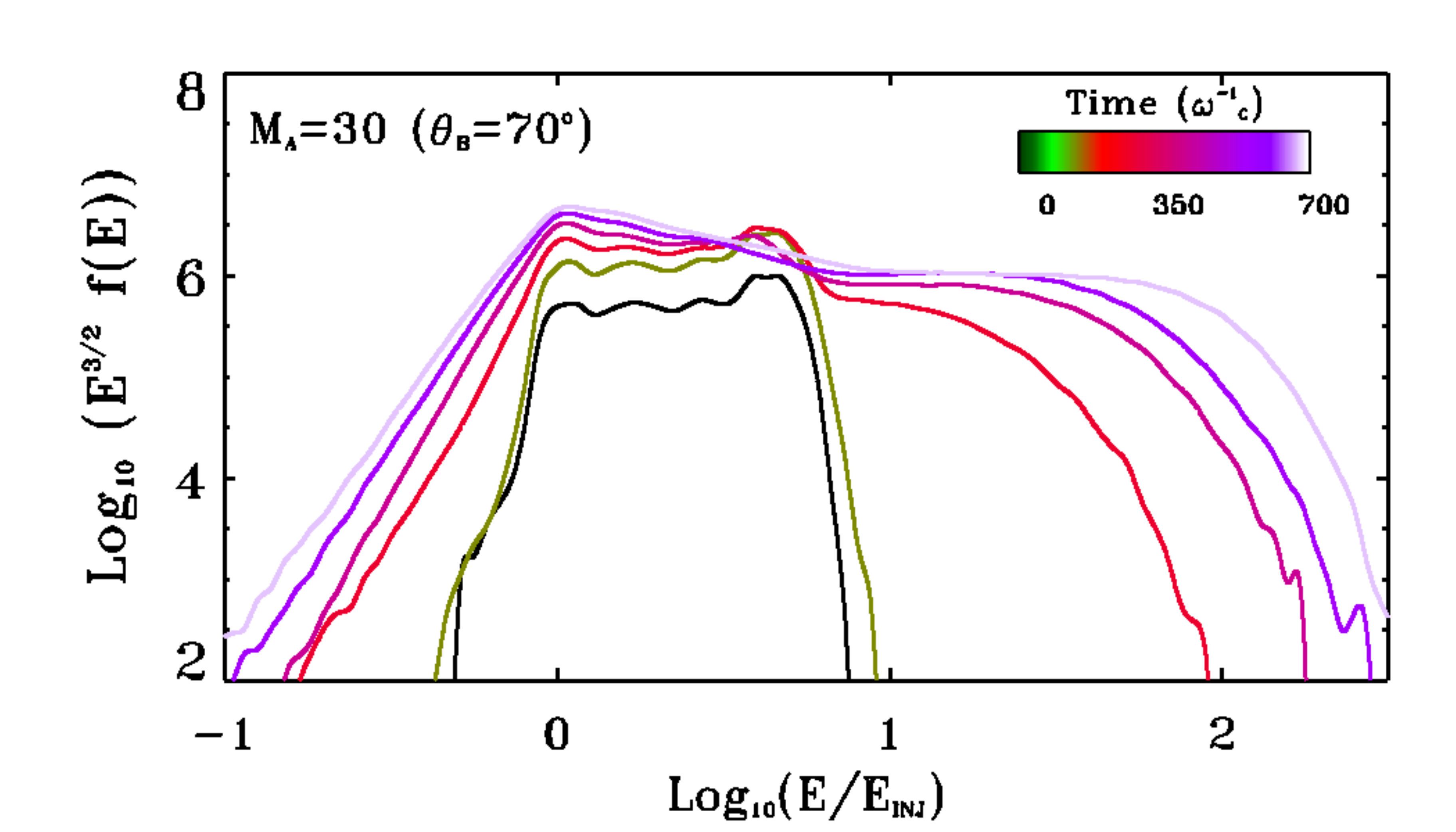}}
\caption{Energy spectra of non-thermal particles injected at energy $E_{\rm inj}/mc^2=4\times 10^{-5}$. The various spectra correspond to simulations displayed in Fig\ref{fig:Current_perp}. 
}
 \label{fig:spectra_perp_M30}
\end{figure}
\begin{figure*}{ht}
\centering
\mbox{
\includegraphics[width=\columnwidth]{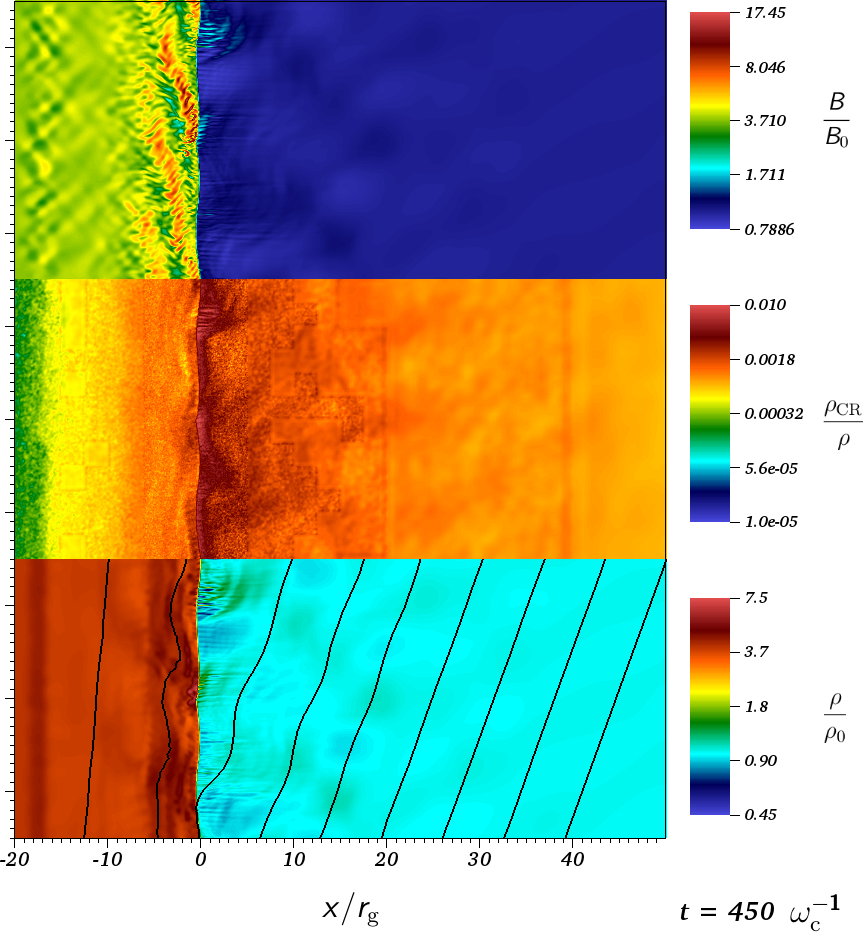}
\includegraphics[width=\columnwidth]{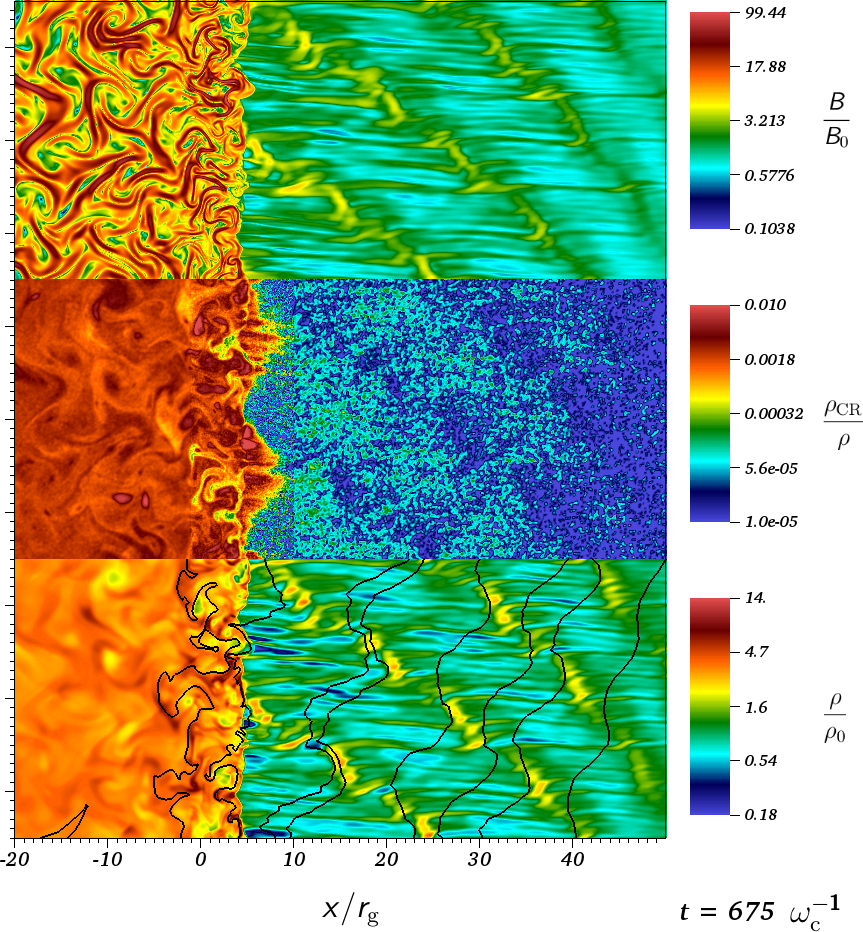}}
\caption{Similar to Figs.~\ref{fig:20deg1} and \ref{fig:parallel4}, but for $M_A=300$ and $\theta_B=70$.
Although the large-scale instabilities in the upstream medium are similar to the $M_A=30$ case, they show a far more elongated pattern as a result of the higher Mach number.
}
 \label{fig:20deg4}
\end{figure*}

The corrugation of the shock affects the magnetic structure in its close vicinity. Indeed we can see on Fig.\ref{fig:Current_perp} that the opening angle of the magnetic field lines increases in some periodic locations along the shock front. Such bending of the field lines creates locally the physical conditions prevailing in near-parallel shocks, hence enabling a large number of particle to enter the upstream medium (see the left panel of Fig.\ref{fig:20deg1} where three channels of particle arise from the downstream medium to flow into the upstream). The massive input of particle triggers magnetic turbulence dominated by the corrugation wavelength. The contrast between early and late spectra is visible in Fig.\ref{fig:Spectrum_perp} where we displayed the magnetic power spectrum of the $B_x$ component for both early and late stages of the simulation. 

In the early phase of the simulation, the upstream spectrum does not show any dominant pattern while in the downstream spectrum, the weak magnetic fluctuations induced by the streaming of particle along the shock front are visible. In the late stages of the simulation, powerful magnetic fluctuations occur in both media whose dominant wavelength corresponds to the shock corrugation wavelength. Once strong turbulence is triggered, supra-thermal particles experience a more chaotic motion regime leading to a diffusive shock acceleration process. 
 
 The energy spectrum of the particles at various times is displayed in Fig.\ref{fig:spectra_perp_M30}. During the early phase, the energy particle distribution is consistent with SDA spectrum where numerous particles get accelerated up to ten times their injection energy while the majority of these particles gets decelerated due to their interaction with the ambient plasma and magnetic field. The energy spectrum remains the same until $t\sim 200\omega_c^{-1}$ when an additional tail appears due to the interaction between the particles and the shock. This tail converges toward a $f(E)\propto E^{-3/2}$ distribution consistent with a diffusive shock acceleration process. It is noteworthy that the maximal energy reached by supra-thermal particles is approximately $300~ E_{\rm inj}$, namely ten times the maximal energy obtained in the parallel shock simulation. Such result is consistent with the combined effect provoked by SDA and DSA.       
\subsubsection{Origin of the shock corrugation}
The corrugation of the shock front appearing in our simulations has not been observed in previous full-PIC or hybrid computations (see Sect.\ref{sec-discussion} for a comparison with previous studies). In order to make sure that this phenomenon is genuinely physical we have performed a series of test and analysis. First of all, computing the supra-thermal particle current density enables us to compare the large-scale wavelength fluctuation occurring downstream of the shock to the predicted dominant mode that would be generated by the NRS instability in the downstream medium. 
Usually, this instability does take place in the upstream region as it displays a larger Alfv\'enic Mach number, hence allowing a wider range of unstable mode to arise. In the context of an highly-oblique shock, the upstream NRS instability is not triggered whereas it can in the downstream medium where the  cosmic-ray current reaches a steady-state more rapidly as no selection through the SDA process is required for downstream particles.
 
The setup of our simulation leads to a downstream medium where the Alfv\'enic Mach number remains larger than one ($M_A\sim 4$). The downstream supra-thermal particle current rapidly decreases away from the shock (see Fig.\ref{fig:Spectrum_perp}). Computing the average value of the current leads to a ratio $J_{\rm part}c/\omega_{\rm pi}B_{\rm down}$ of the order of $0.018$. Such a ratio is consistent with a dominant unstable mode whose wavelength is $\lambda\sim 720 c/\omega_{\rm pi}=8 r_g$. This is consistent with the wavelength observed in the magnetic fluctuation (see Fig.\ref{fig:profile_B_perp}). On the other hand the characteristic growth timescale in such conditions is $t_{\rm gr,max}=\omega_{\rm pi}B_{\rm down}/J_{\rm part}c~\omega_c^{-1}\simeq 50 \omega_c^{-1}$. The growth timescale is quite large but consistent with the fact that we do see the rise of the corrugating magnetic fluctuation after at least $200~\omega_c^{-1}$ (see Fig.\ref{fig:profile_B_perp}). 

At this point we have performed a series of tests in order to ensure that the corrugation of the shock does not result from any numerical artifact. To that end, we have run a simulation where we have increased the $y$ extension of the computational domain by a non-integer number, namely $2.5$. In addition to that modification, we have also turn off the AMR procedure and set the resolution of the uniform grid to the highest spatial resolution used in AMR computations. We have recover, using the same initial conditions, exactly the same corrugation phenomenon with the same wavelength and particle spectrum. Such results comforted us that the corrugation of the shock is not induced by the setup of the computation domain nor its spatial extension. We then turn our attention to the injection procedure of the particles. We have done another series of runs where we have modified the charge of the particle or their mass, hence modifying the Larmor radius of the injected particles. It turns out that in computation 
where we only changed the mass but not the charge of the particles, we recover the same corrugation pattern with the same wavelength. On the other hand, when changing the charge of the particle while leaving untouched their mass, we recover the corrugating pattern but with a different wavelength proportional to the change in the electrical charge of the particle. These computations then prove that the origin of the corrugating pattern is only dependent on the supra-thermal particle current and not on the particle Larmor radius. Such result rules out the resonant streaming instability and proves that the NRS instability is very likely the mechanism at the origin of the corrugation of the shock.

Finally we also mention that no particle acceleration and no shock corrugation have been observed in a strictly perpendicular shock at times as long as $1000~\omega_{\rm c}^{-1}$.

\subsection{Higher Alfv{\'e}nic Mach number simulations}
Fig.~\ref{fig:20deg4} shows what happens when we combine a higher Alfv{\'e}nic Mach number $M_A=300$) shock with a magnetic field that makes a 70 degree angle with the direction of motion. 
Initially, particle acceleration occurs in a similar fashion as for the $M_A=30$ case, namely through a SDA mechanism. We eventually also observe a corrugating shock front on the same timescale than in the $M_A=30$ simulation. It is important to mention here that in order to have an upstream Mach number $M_A=30$ we increased the velocity of the shock by a factor ten. Accordingly, we have increased the injected particle velocity $v_{\rm inj}$ by a factor ten so that the fiducial length and time, $\omega_c^{-1}\propto v_{\rm inj}^{-1}$ and $r_g\propto v_{\rm inj}$ are also changed. Keeping the same injection recipe leads to a similar downstream supra-thermal current that is still in agreement with the observed corrugation wavelength. This result suggests that the NRS instability is indeed the mechanism at work in the corrugation of the shock front.

However, we do observe some effect of the Mach number on the upstream medium.  
On top of the patterns that we found in the  $M_A=30$ case, we see a second instability that creates filament along the incoming flow. We do believe that this is the result of the high flow speed, which prohibits any significant motion perpendicular to the direction of the flow hence dislocating the transverse filamentation occurring once the shock is fully corrugated (see right panel of Fig.\ref{fig:20deg4}). 

\begin{figure}
\centering
\mbox{
\includegraphics[width=\columnwidth]{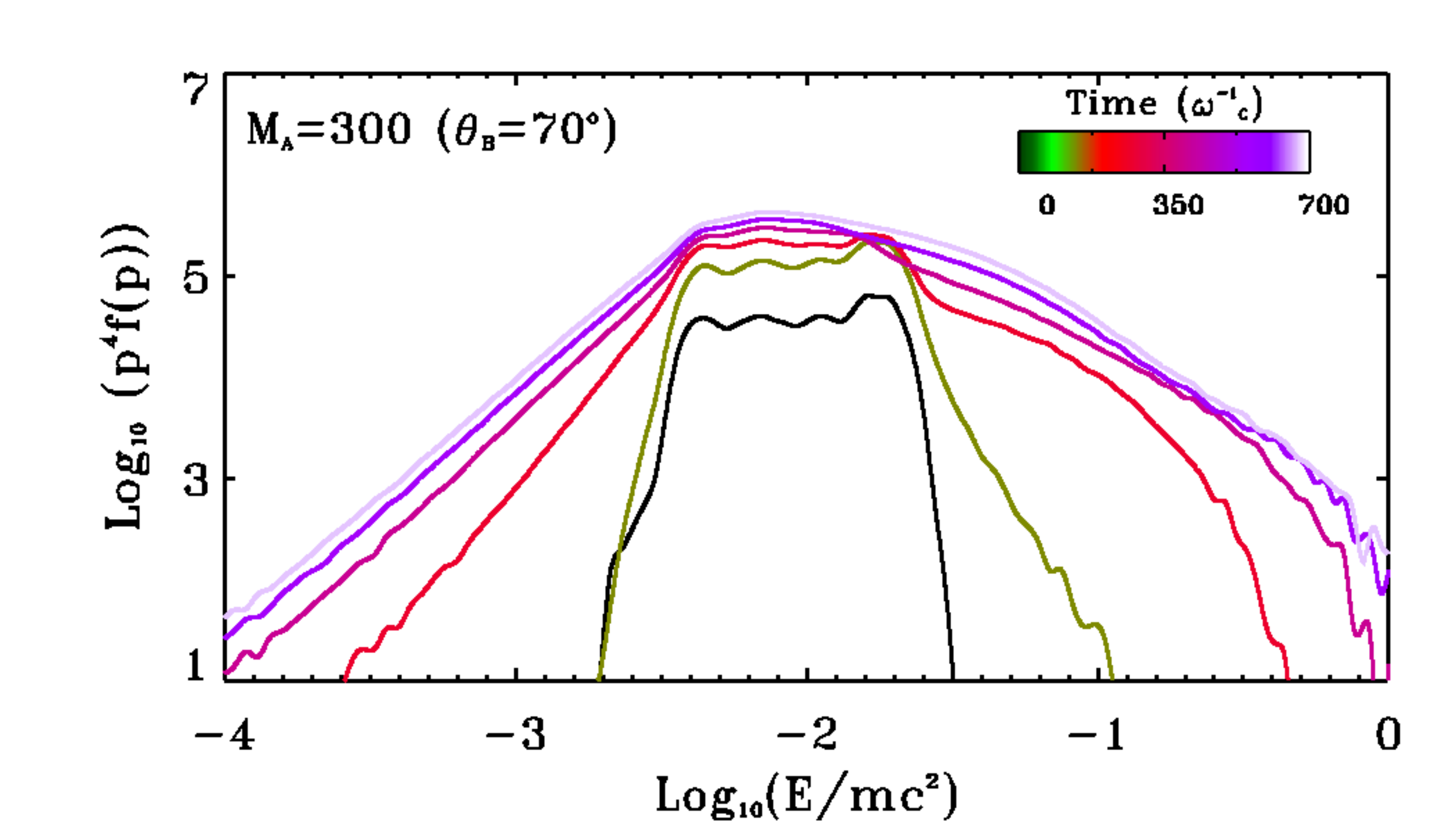}}
\caption{Energy spectra of non-thermal particles injected at energy $E_{\rm inj}/mc^2=4\times 10^{-3}$ near a $M_A=300$ shock. The various spectra correspond to simulation displayed in Fig.\ref{fig:20deg4}. It is noteworthy that some supra-thermal being accelerated through a combined SDA-DSA process are able to reach the relativistic regime (with Lorentz factors of the order of a few units).} 
 \label{fig:spectra_para_M301_6}
\end{figure}
The energy spectrum  of the supra-thermal particles in the $M_A=300$ simulation is displayed on Fig.\ref{fig:spectra_para_M301_6}. The shape of the spectrum is nearly identical to the $M_A=30$ case, namely an initial pile-up of accelerated particle through SDA completed afterwards by a higher energy tail associated with a DSA process. It is noteworthy that in this simulation the injected non-relativistic particles are able to reach a (moderate) relativistic regime as the Larmor radius of the most energetic particles are of the order of a few units.
\section{Discussion}
\label{sec-discussion}
Our parallel shock model succeeds in reproducing earlier results obtained with both a di-hybrid code \citep[e.g.]{Caprioli14a} and a combined PIC-MHD code \citep{Baietal:2015}. They also confirm the theoretical predictions for NRS instabilities by \citet{Bell04,Bell05}. 
While performing these simulations we also tested the effect of varying the angle between the flow and the magnetic field and find results similar to those of e.g. \citep{Caprioli14a} for small angles (up to approximately 45 degrees.) 
In contrast, the oblique shock model shows a new, unexpected result, which deviates considerably from the results obtained previously by \citet{Caprioli14a}. It is noteworthy that such simulations have never been done using a PI[MHD]C code so direct comparison is impossible. In Section \ref{sec-Bfieldangle} we have discussed the mechanisms that lead to the formation of instabilities in both the upstream and downstream medium when the magnetic field makes a large angle with the direction of the flow. 
We should note here that when we increase the angle even further beyond $70^o$ this effect disappears since the component of the particle velocity parallel to the flow becomes smaller than the drift velocity if the thermal plasma, which eliminates the upstream current. 
The question remains as to why this result was not obtained with an hybrid code. By comparing our simulations with these earlier models we come to the conclusion that there are two factors that contribute to the different outcome. 
\begin{enumerate}
\item The number of particles. Although the absolute number of particles used in hybrid codes is far larger than what we use for the PI[MHD]C simulations, only a small fraction of the particles in hybrid simulations actually are supra-thermal. Combined with the fact that for an oblique magnetic field only a fraction of the non-thermal particles is able to move upstream because of their pitch angle, we conclude that the hybrid simulations lacked the necessary number of particles to create the current that causes the disturbance of the upstream magnetic field. 
\item The size of the physical domain. The size of the simulated domain in the direction perpendicular to the shock-normal is six times larger in our model than in the original hybrid simulation. This allows us to capture the long wavelength which corrugates the shock and sets in motion the process that disturbs the upstream magnetic field. The smaller domain used by the hybrid simulation makes this impossible. 
We have tested our code with a physical domain similar to that used by \citet{Caprioli14a} and find that indeed the long wavelength never appears and so is the magnetic field amplification. 
\end{enumerate}
These points demonstrate the opportunity that the PI[MHD]C method affords us. Because it is designed primarily to simulate large scale structures, is less computationally expensive than PIC or hybrid methods, and enables us to run the simulation in the frame of reference of the shock, it allows us to simulate a large volume of space,  using a larger number of supra-thermal particles, which in this particular case proved to be necessary. 

2D3V PIC simulations of perpendicular and oblique shock configurations have been performed by \citet{Kato10} and  \citet{Wieland16} respectively. These studies verify that fast Alfv\'enic Mach (low magnetization) number shocks are mediated by the Weibel instability. These studies only report on electron thermalization but find evidences for the development of a proton non-thermal tail possibly connected with a shock surfing process in the perpendicular shock configuration. We confirm with \citet{Caprioli14a} that protons are accelerated at oblique shocks with an angle $\theta_{\rm B} = 45^o$. However, we argue that particle acceleration also proceeds at larger angles through the two-steps instability process described in section \ref{sec-Bfieldangle} as identified by \citet{Reville13}. But we do not observe any acceleration at perpendicular shocks at the timescales explored in the simulations.

Ideally, we would like to go for a long PIC or an hybrid simulation to be run in order to verify our results. However, such a simulation would require considerable computational resources. The simulation box would have to be large enough along the axis parallel to the flow to be able to follow the shock for a time frame of at least 400\,$\omega_c^{-1}$ to capture both the formation of the long wavelength pattern in the upstream medium and the response of the shock and downstream medium to this development. Along the perpendicular axis it would need to be large enough to capture the long wavelength. Finally, the number of particles would have to be increased, not only to compensate for the larger volume, but also to provide sufficient supra-thermal particles to reproduce the upstream current. 

It should be kept in mind that we are using a  2-D model to investigate what is fundamentally a 3-D problem. We have chosen to do so in order to limit the computational cost of our simulations and to reproduce the existing 2-D models for purpose of comparison. The non-resonant streaming instability is an inherently 3-D phenomenon. Reducing it to 2-D is likely to alter its growth and encourages the formation of filaments in the upstream medium, rather than a true three-dimensional structure.

\section{Conclusions}
\label{sec-conclusions}
Our simulations prove that our code can reliably reproduce earlier work \citep{Baietal:2015} as well as analytically predicted effects. We recover the ignite of the NRS in the parallel shock configuration which further produces a filamentation of the upstream plasma. Particle acceleration takes place due to the DSA mechanism. In the oblique shock configuration we also find the development of magnetic field amplification as well as particle acceleration. But these occur in a two-steps process: first particle are accelerated by SDA and gain energy up to 9 times the injection energy. This induces in turn a CR current able to trigger the NRS instability downstream which finally leads to a strong shock corrugation. At locations along the shock front where the magnetic field is sufficiently perturbed to retrieve a quasi-parallel configuration bunches of CR are able to explore the upstream medium and trigger a NRS instability and a filamentation of the medium as in the parallel shock case. Once the NRS instability 
is onset particles start to gain energy under the effect of DSA. Maximum particle energies at the same timescales are found to be higher in the oblique configuration with respect to parallel shock case. 

The simulations prove that for any kind of astrophysical shock, the particle acceleration and magnetic field amplification play an important part in determining the characteristics of the shock. We stop our simulations at 675 gyro times. 
Assuming that the magnetic field equals $5\times10^{-6}$\,G (Typical for interstellar magnetic fields), this means that for the $M_A\,=\,30$ case, approximately $1.5\times 10^4$ seconds have passed. For the $M_A\,=\,300$ case, it is $1.5\times 10^5$ seconds. 
Both times are short compared to the typical lifetime of circumstellar and interstellar shocks. Clearly, the presence of non-thermal particles start to influence the behavior of a shock almost as soon as it is formed.

However, this study is quite preliminary in the sense it is 2D3V in dimensions. 3D simulations are mandatory to explore further the efficiency of the NRS instability. At present only non-thermal proton transport has been considered, our next study will also include electrons. For this we are adding a module calculating radiative losses (synchrotron, inverse Compton). Our study needs also to be extended to the case of relativistic flows. Both developments are foreseen to be delivered in a forth coming study.

\section*{Acknowledgements}
We are grateful for fruitful regular discussions with: A.R. Bell, C.A. Bret, A. Bykov, M. Dieckmann, E. D'humi\`eres, M. Grech, L. Gremillet, R. Keppens, B. Lemb\`ege, M. Lemoine, G.Pelletier, I. Plotnikov, V. Tikhonchuk. This work is supported by the ANR-14-CE33-0019 MACH project.
We would like to thank D. Caprioli and B. Reville for helpful comments. We also thank our anonymous referee for helpful comments and suggestions. 
This work acknowledges financial support from the UnivEarthS Labex program at Sorbonne Paris Cit\'e  (ANR-10-LABX-0023 and ANR-11-IDEX-0005-02).  This work was granted access to HPC resources of CINES under the allocation A0020410126 made by GENCI (Grand Equipement National de
Calcul Intensif). 

\bibliographystyle{yahapj}
\bibliography{vanmarle_biblio.bib}

\appendix
\section{Animations}

\begin{figure}
\centering
\mbox{
\includegraphics[width=\columnwidth]{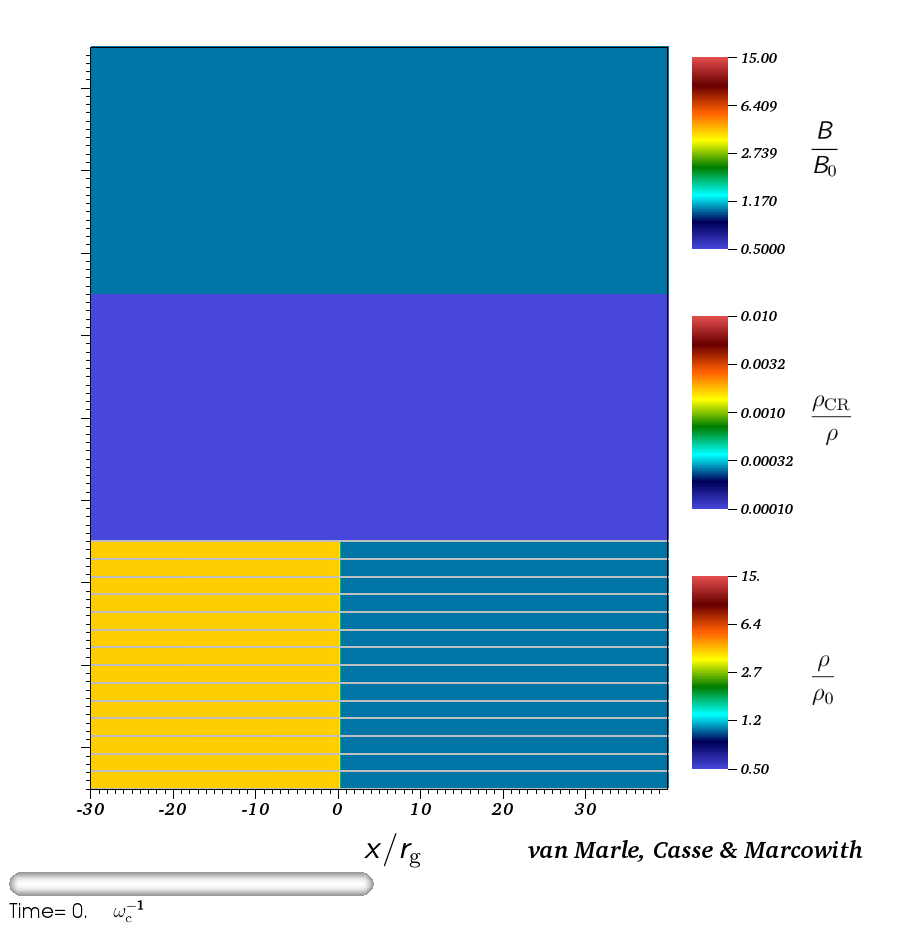}}
\caption{In this animation of a parallel shock with $M_A=30$ we show the evolution of the magnetic field strength relative to the original magnetic field (top), non-thermal particle charge density relative to the thermal gas density (middle), and thermal gas mass density relative to the upstream density at the start of the simulation, combined with the magnetic field stream lines (bottom). The gas is streaming through the shock from right to left. Over time supra-thermal particles are introduced at the shock, causing instabilities in the thermal gas.}
 \label{fig:animation1}
\end{figure}

\begin{figure}
\centering
\mbox{
\includegraphics[width=\columnwidth]{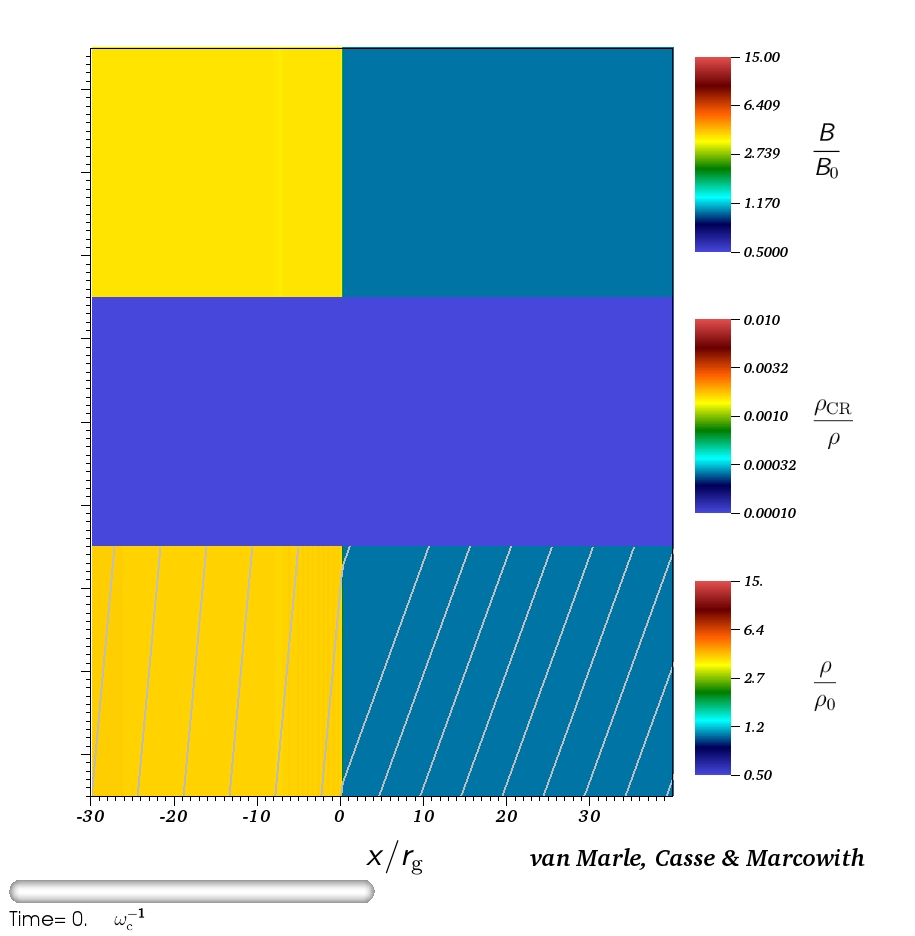}}
\caption{Similar to Fig.~\ref{fig:animation1}, but for the simulation with $M_A=30$ and $\theta_B=70^{\rm o}$. The animation clearly demonstrates the appearance of the long wavelength pattern in the upstream gas.}
 \label{fig:animation2}
\end{figure}


\bsp	
\label{lastpage}
\end{document}